\documentclass[aps,prl,reprint,amsmath,superscriptaddress]{revtex4-2}

\usepackage[colorlinks,
	linkcolor=blue,
	anchorcolor=blue,
	citecolor=blue,
	urlcolor=blue	
]{hyperref}

\usepackage{graphicx}
\usepackage{orcidlink}
\usepackage{lineno}

\makeatletter
\newcommand{\rev}[1]{#1}  
\makeatother

\begin{document}
\title{Cavity Controls Core-to-Core Resonant Inelastic X-ray Scattering}

\author{S.-X. Wang\orcidlink{0000-0002-6305-3762}}
\affiliation{Department of Modern Physics, University of Science and Technology of China, Hefei, Anhui 230026, China}
\affiliation{I. Physikalisches Institut, Justus-Liebig-Universität Gießen und 
Helmholtz Forschungsakademie Hessen für FAIR (HFHF) GSI Helmholtzzentrum für 
Schwerionenforschung Campus Gießen, Heinrich-Buff-Ring 16, 35392 Gießen, 
Germany}

\author{Z.-Q. Zhao}
\author{X.-Y. Wang}
\author{T.-J. Li\orcidlink{0000-0002-4391-5257}}
\author{Y. Su}
\affiliation{Department of Modern Physics, University of Science and Technology of China, Hefei, Anhui 230026, China}

\author{Y. Uemura\orcidlink{0000-0003-3164-7168}}
\author{F. Alves Lima\orcidlink{0000-0001-8106-2892}}
\affiliation{European XFEL, 22869 Schenefeld, Germany}

\author{A. Khadiev\orcidlink{0000-0001-7577-2855}}
\author{B.-H. Wang\orcidlink{0000-0002-1223-503X}}
\affiliation{Deutsches Elektronen-Synchrotron DESY, 22607 Hamburg, Germany}

\author{J. M. Ablett\orcidlink{0000-0003-2887-2903}}
\affiliation{Synchrotron SOLEIL, L’Orme des Merisiers, Départementale 128, 91190 Saint-Aubin, France}

\author{J-P. Rueff\orcidlink{0000-0003-3594-918X}}
\affiliation{Synchrotron SOLEIL, L’Orme des Merisiers, Départementale 128, 91190 Saint-Aubin, France}
\affiliation{Sorbonne Université, CNRS, Laboratoire de Chimie Physique – Matière et Rayonnement, LCPMR, F-75005 Paris, France}

\author{H.-C. Wang}
\author{O.J.L. Fox\orcidlink{0000-0001-5224-7062}}
\affiliation{Diamond Light Source, Harwell Science and Innovation Campus, Didcot, Oxfordshire, OX11 0DE, United Kingdom}

\author{W.-B. Li}
\affiliation{MOE Key Laboratory of Advanced Micro-Structured Materials, Institute of Precision Optical Engineering 
(IPOE), School of Physics Science and Engineering, Tongji University, Shanghai 200092, China}

\author{L.-F. Zhu\orcidlink{0000-0002-5771-0471}}
\affiliation{Department of Modern Physics, University of Science and Technology of China, Hefei, Anhui 230026, China}

\author{X.-C. Huang\orcidlink{0000-0002-9140-6369}}\email[Corresponding author: ]{xinchao.huang@xfel.eu}
\affiliation{European XFEL, 22869 Schenefeld, Germany}

\date{\today}

\begin{abstract}

    X-ray cavity quantum optics with inner-shell transitions has been hindered 
    by the overlap between resonant and continuum states. Here, we report the 
    first experimental demonstration of cavity-controlled core-to-core resonant 
    inelastic x-ray scattering (RIXS). We eliminate the effects of the 
    absorption edge by monitoring the RIXS profile, thereby resolving the 
    resonant state from the overlapping continuum. We observe distinct 
    cavity-induced energy shifts and cavity-enhanced decay rates in the $2p3d$ 
    RIXS spectra of WSi$_{2}$. These effects, manifesting as stretched or 
    shifted profiles in the RIXS planes, enable novel spectroscopic 
    applications by cavity-controlled core-hole states. Our results establish 
    core-to-core RIXS as a powerful tool for manipulating inner-shell dynamics 
    in x-ray cavities, offering new avenues for integrating quantum optical 
    effects with x-ray spectroscopy.

\end{abstract}

\maketitle

\paragraph{Introduction}X-ray quantum optics has become a rapidly evolving research field \cite{adams2013, kuznetsova2017, wong2021, wang2024acta} driven by advancements in x-ray sources \cite{Bostedt2011, Einfeld2014} and sample fabrication. In particular, thin-film planar cavities serve as a versatile platform for exploring quantum optical effects in the hard x-ray regime, enabling the observation of fundamental phenomena such as collective Lamb shift, superradiance, and strong light-matter coupling \cite{röhlsberger2020, röhlsberger2021}. Cavities can modify the photonic density of states, thereby controlling the dynamics of excited states \cite{raimond2001}. Initially, thin-film cavities were employed for Mössbauer nuclear transitions, exploiting their narrow linewidths and long coherence times \cite{rohlsberger2005, rohlsberger2010, rohlsberger2012, heeg2015fano, heeg2015slow, haber2017, huang2017, velten2024}. More recently, they have been extended to manipulate inner-shell electronic transitions, which exhibit intriguing core-hole dynamics \cite{haber2019, huang2021, vassholz2021, gu2021, ma2022, huang2024}. Notable advancements, including spectral manipulation \cite{haber2019}, novel Fano profile \cite{ma2022}, core-hole lifetime controlling \cite{huang2021}, and core polaritons \cite{gu2021}, have been successfully demonstrated, among others. Differing from the well-defined narrow transitions of Mössbauer nuclei \cite{röhlsberger2020, röhlsberger2021} or other quantum systems in the visible and microwave regimes \cite{blatt2012, blais2021}, inner-shell transitions involve unique intermediate core-hole states characterized by multiple decay pathways, which in turn enables various modern x-ray spectroscopy techniques \cite{gelmukhanov2021, wang2024acta}. 

Core-hole states lie at the heart of core-level x-ray spectroscopies 
\cite{groot2008}. \rev{For example, core-hole dynamics play a crucial role in 
resonant inelastic x-ray scattering (RIXS), in which an incident x-ray excites 
a core electron to an unoccupied state and the system subsequently decays by 
emitting a photon as another electron fills the hole} \cite{kotani2001, 
ament2011}. \rev{RIXS serves as a key probe  of atomic, molecular, and 
condensed matter systems,} across physics, chemistry, and biology. \rev{These 
dynamics} shape the temporal evolution of the scattering amplitude 
\cite{gelmukhanov1999}, control intensity ratios among elementary excitations 
\cite{ament2011, ghiringhelli2005}, and modulate transition strengths between 
bound states \cite{baker2017}, among other effects. The core-hole state has 
traditionally been regarded as an intrinsic property of matter, largely due to 
the lack of control over its decay channels. Recent advances, however, have 
demonstrated that both the \rev{decay rates} and transition energy of 
\rev{inner-shell transitions} can be tuned using x-ray cavities 
\cite{haber2019, huang2021, huang2024}. Integrating such cavity control with 
RIXS introduces new opportunities \cite{gelmukhanov2021, wang2024acta} for 
manipulating core-level interactions and expanding the scope of x-ray 
spectroscopic techniques.

\rev{In contrast to the well-isolated transitions available in Mössbauer 
nuclei, spectral} overlap between resonant and continuum states 
\cite{haber2019, huang2021, ma2022, huang2024, wang2024acta} has so far limited 
the observation of quantum optical effects \rev{in inner-shell transitions 
}\cite{rohlsberger2012, heeg2013prl, kong2016, haber2017}. RIXS offers a 
promising approach to circumvent this limitation by exploiting resonant Raman 
processes to selectively probe resonant transitions \cite{gelmukhanov2021}. 
Realizing x-ray cavity control over RIXS would therefore mark a significant 
step forward, enabling not only advanced spectroscopic applications but also 
new regimes of x-ray quantum optics. \rev{However, }RIXS experiments face 
substantial challenges for signal detection due to their low scattering cross 
sections \cite{schulke2007}. Achieving the high energy resolution needed to 
resolve subtle spectral features requires highly monochromatic x-rays and 
crystal analyzers with a relatively large radius of curvature, making RIXS a 
photon-hungry technique \cite{kotani2001, ament2011}. The incorporation of 
x-ray cavities (with ultra-thin atomic layers) imposes further constraints, 
requiring highly collimated beams to enhance cavity effects \cite{heeg2013pra, 
heeg2015pra, kong2020, lentrodt2020prr, huang2024}, which further limits photon 
throughput. Core-to-core RIXS \rev{(in which another core electron fills a core 
hole created by x-ray absorption)}, often termed resonant x-ray 
Raman scattering \cite{ament2011} or resonant x-ray emission (RXES) 
\cite{groot2001high}, offers comparatively higher scattering intensity owing to 
its two-step, dipole-allowed nature. This makes it the most promising approach 
for initial demonstration aimed at merging quantum optics with RIXS. 
Nonetheless, practical challenges remain, particularly in realizing 
grazing-incidence geometries required for efficient cavity coupling in the 
x-ray regime.

In this work, we report the first experimental demonstration of cavity control 
over core-to-core RIXS using a thin-film planar cavity. \rev{Guided by quantum 
Green's function theory \cite{kong2020, lentrodt2020prr, huang2024}, the cavity 
was designed to strongly modify the core-hole state. To probe this effect we 
collected two-dimensional RIXS planes (maps of x-ray emission intensity versus 
incident and emitted photon energies), using a high-resolution 
energy-dispersive von Hamos (VH) spectrometer. Multiple crystal analyzers were 
used to boost the detection solid angle. We also benchmarked the setup by 
measuring the RIXS plane at a large incident angle in the absence of cavity 
effects. Subsequent measurements at two distinct cavity detunings revealed 
hallmark quantum optical effects: a cavity-induced energy shift (CIS) and a 
cavity-enhanced decay rate (CER), both clearly observed in the RIXS plane as 
energy and intensity modulations. These observations open the door to advanced 
cavity-coupled x-ray spectroscopy based on core-to-core RIXS, including high 
energy resolution off-resonant spectroscopy (HEROS) \cite{blachucki2014} and 
high energy resolution fluorescence detected (HERFD) absorption spectroscopy 
\cite{bauer2014}. Moreover, by isolating the resonant Raman feature at constant 
energy transfer \cite{baker2017, groot2005}, we circumvent the effects of 
transition-edge that typically obscure inner-shell transitions in hard x-ray 
quantum optics \cite{haber2019, huang2021, ma2022, huang2024}. }

\begin{figure}[!htbp]
    \centering
	\includegraphics[width=\linewidth]{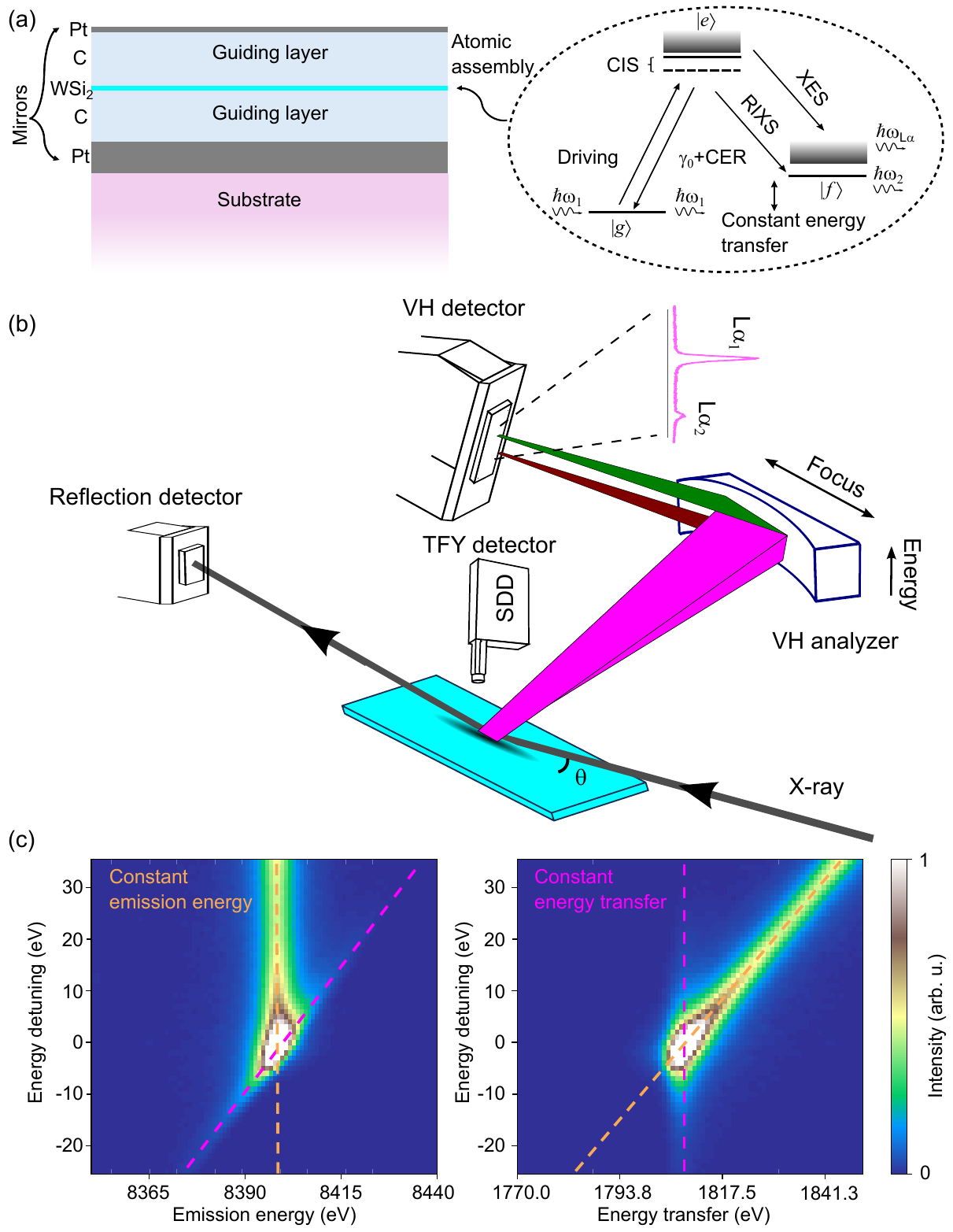}
	\caption{Experimental scheme. (a) Cavity structure and RIXS energy-level 
	diagram. Left: multilayer cavity composed of Pt mirrors, carbon (C) guiding 
	layers, and a WSi$_2$ atomic ensemble. Right: simplified energy-level 
	diagram for the core-to-core RIXS. \rev{ The driving field resonantly 
	excites the system to an intermediate core-hole state, which subsequently 
	decays to the final state by emitting a characteristic x-ray. During the 
	process, cavity effects induce an energy 
	shift and an enhanced decay rate to the intermediate state.} (b) 
	Measurement geometry. The thin-film cavity sample is aligned at grazing 
	incidence, resulting in a large footprint on the surface. A von Hamos (VH) 
	spectrometer (energy resolution $\sim$1 eV) disperses and focuses the 
	emitted x-rays onto a 2D detector, recording energy-resolved spectral 
	images. A silicon drift detector (energy resolution $\sim$200 eV) collects 
	the total fluorescence yield in the vertical direction, while a downstream 
	2D detector monitors cavity reflectivity. (c) 2D RIXS planes shown as a 
	function of emission energy (left) or energy transfer (right). A constant 
	background, estimated from off-resonant spectral tails, has been 
	subtracted. The incident angle (8.7 mrad) was set to be above the critical 
	angle, a condition under which the x-ray can penetrate the sample; thus, 
	the cavity effect is minimized. Pink dashed lines indicate the resonant 
	Raman peaks at constant energy transfer, and yellow dashed lines mark the 
	constant emission energy.}\label{fig1}
\end{figure}

\paragraph{Cavity effects on RIXS}The excitation and de-excitation processes in 
RIXS are non-separable, which can be described by the Kramers-Heisenberg 
formula \cite{ament2011}:
\begin{equation}\label{kheq}
    \begin{split}
        \frac{{{d^2}\sigma }}{{d\Omega d\omega }} 
        &= {\sum\limits_f {\left| {\sum\limits_e {\frac{{\left\langle f 
        \right|\vec{D'} \left| e \right\rangle \left\langle e \right|\vec{D} 
        \left| g \right\rangle }}{{{E_g} + {\omega _1} - {E_e} - \delta_c + 
        \frac{{i\gamma }}{2}}}} } \right|} ^2} \\
        & \times \delta \left( {{E_g} + {\omega _1} - {E_f} - {\omega _2}} 
        \right)  
    \end{split}
\end{equation}
where $\omega_{1,2}$ are the energies of the incident and scattered photons, 
respectively, and $E_{g,e,f}$ denote the energies of ground 
$\left|g\right\rangle$, intermediate (core-hole) $\left| e \right\rangle$ and 
final $\left| f \right\rangle$ states. The dipole operators $\vec{D}$ and 
$\vec{D'}$ govern the absorption and emission steps, respectively. 
Momentum-dependent effects are neglected, as core-level states are typically 
non-dispersive. The parameter $\delta_c$ represents the CIS while $\gamma = 
\gamma_0 + \gamma_c$ includes both the natural decay rate $\gamma_0$ and the 
CER $\gamma_c$. \rev{Figure \ref{fig1}(a) schematically illustrates the 
core-to-core RIXS process. In this work, an incident x-ray of energy $\omega_1$ 
resonantly excites a 2\emph{p} electron into an unoccupied state and a 
3\emph{d} electron subsequently fills the created core-hole, emitting an x-ray 
of energy $\omega_2$, corresponding to L$\alpha$ line for WSi$_{2}$. The energy 
transfer $\omega_1 - \omega_2$ reflects the excitation energy of the final 
state. For resonant scattering, 
the energy transfer remains constant as $\omega_1$ is tuned across the 
absorption edge, giving rise to a vertical feature in the RIXS plane. Above the 
ionization threshold, however, the excited electron enters the continuum, and 
the subsequent decay emits x-rays with a nearly fixed energy, yielding a 
diagonal feature at a constant emission energy. Figure \ref{fig1}(c) presents a 
representative $2p3d$ RIXS plane of WSi$_2$ (without cavity effects), plotted 
as a function of both emitted photon energy (left panel, often termed RXES) and 
energy transfer (right panel, RIXS). In the latter representation, transitions 
to bound and continuum states are clearly distinguished—an advantage not 
achievable with conventional reflectivity or total-fluorescence-yield (TFY) 
measurements \cite{haber2019, huang2021, ma2022, supp2024}.}

\begin{figure}[htbp]
	\centering
	\includegraphics[width=0.8\linewidth]{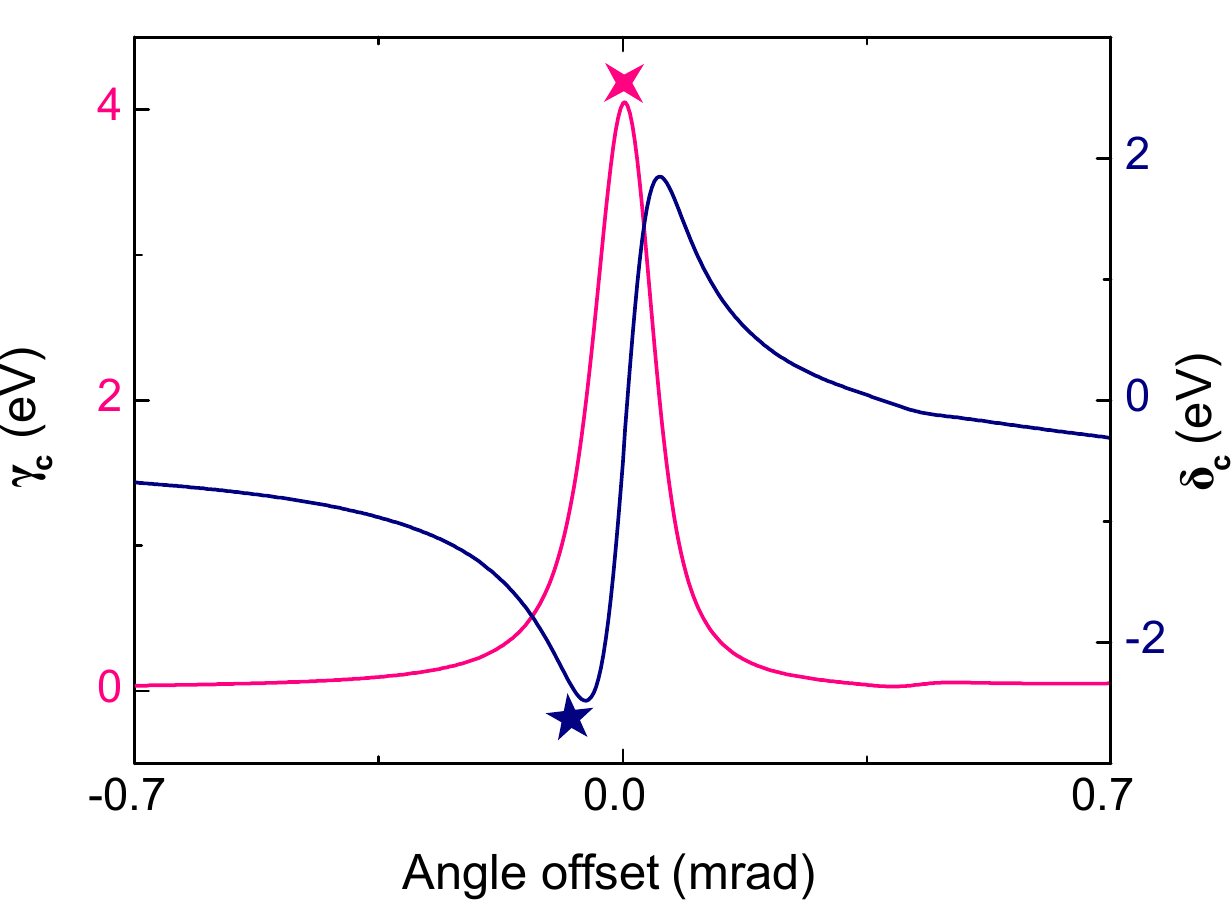}
	\caption{The CER and CIS as a function of angle offset calculated by the quantum Green's function model. An angle divergence of 30 $\mu$rad is convolved with the theoretical calculation. The pink cross and blue star indicate the specific angle offsets used in the experimental measurements, corresponding to the cavity-mode angle and an angle offset of $70~\mu$rad, respectively.
    \label{fig2} }
\end{figure}

The interaction of the x-ray cavity with the inner-shell resonances can be well described by the recently developed quantum optical model \cite{huang2024}. In particular, the collective spin-exchange $J$ (associated with $\delta_c$) and decay rate $\Gamma$ (associated with $\gamma_c$) are expressed as:
\begin{eqnarray}
    \label{jlleq}
	J &=& \frac{N}{A} {\mu_{0}\omega^{2}} 
	\mathbf{d}^{*} \cdot {\mathrm{Re}}[\mathbf{G}(z, z^{\prime}, \omega)] \cdot \mathbf{d}, \\
    \label{gammalleq}
	\Gamma &=& \frac{2N}{A} {\mu_{0}\omega^{2}} 
	\mathbf{d}^{*} \cdot {\mathrm{Im}}[\mathbf{G}(z, z^{\prime}, \omega)] \cdot \mathbf{d}.
\end{eqnarray}
where $\frac{N}{A}$ is the atomic number area density, $\mathbf{d}$ is the 
dipole matrix element, and $\mathbf{G}(z, z^{\prime}, \omega)$ is the 
normalized Green's function which can be calculated using a transfer matrix or 
Parratt's formalism \cite{huang2024}. Within the x-ray cavity, $\mathbf{G}(z, 
z^{\prime}, \omega)$ strongly depends on the cavity detuning, particularly the 
angular offset ($\Delta \theta$) from the first-order cavity mode. Using Eqs. 
\ref{jlleq} and \ref{gammalleq}, Fig. \ref{fig2} visualizes the control 
mechanism of the core-hole states \rev{under the cavity used in this work, 
consisting of 2.4-nm Pt / 20.4-nm C / 2.9-nm WSi$_{2}$ / 20.0-nm C / 14.0-nm 
Pt, deposited on a silicon wafer. WSi$_{2}$ serves as the resonant atomic layer 
due to its strong $2p-5d$ dipole-allowed transition, and the RIXS measurement 
with bulk samples shows no additional splittings due to crystal-field and 
spin-orbit coupling \cite{zhao2025}. The performance and the structure of the 
cavity sample were characterized at the B16 Test Beamline at the Diamond Light 
Source (as detailed in \cite{supp2024}).} The maximum CER appears at the 
cavity-mode angle where the CIS is relatively small, while the maximum CIS 
occurs at  $\Delta \theta \sim 70 ~\mu$rad where CER is relatively small. These 
two characteristic cavity detunings were selected for experimental 
investigation and are marked by the pink cross and the blue star in Fig. 
\ref{fig2}.

\paragraph{Scheme of RIXS measurement} 
\rev{The x-ray thin-film x-ray planar cavity is operated at grazing-incidence 
angles (typically a few milliradians) to maintain high reflectivity of the 
mirror layers, resulting in an elongated beam footprint ($>$10 mm) and a 
line-like emission region on the sample. Previous studies recorded x-ray 
reflectivity \cite{rohlsberger2005, rohlsberger2010, rohlsberger2012, 
heeg2015fano, heeg2015slow, haber2017, huang2017, velten2024} and TFY 
absorption spectra \cite{haber2019, huang2021, ma2022} with relatively low 
energy resolution, for which an extended footprint does not degrade their 
measured quality at a fixed angle. In contrast, RIXS measurements require high 
energy resolution, while the line-like emission source at grazing incidence 
causes defocusing and resolution loss in conventional analyzer geometries 
(e.g., Johann \cite{johann1931} or DuMond \cite{dumond1947} types). To minimize 
this limitation, we employed a von Hamos (VH) spectrometer \cite{hamos1933, 
wach2020} to record the RIXS spectra, as illustrated in Fig.~\ref{fig1}(b). The 
VH configuration uses a cylindrically bent crystal that disperses photon 
energies along the axial axis 
via Bragg diffraction while simultaneously focusing the signal along the bent 
direction. By aligning the elongated emission line with the cylindrical 
focusing axis,  the VH spectrometer decouples the large beam footprint from the 
dispersion plane as much as possible, thereby preserving the energy resolution  
(see \cite{supp2024} for details).} The inset of Fig.~\ref{fig1}(b) shows 
well-resolved L$\alpha_1$ ($3d_{5/2} \rightarrow2p_{3/2}$) and L$\alpha_2$ 
($3d_{3/2} \rightarrow2p_{3/2}$) emission lines, where the dominant L$\alpha_1$ 
emission was selected to record the RIXS spectra by scanning the incident x-ray 
energy. The measurements were performed at the GALAXIES beamline of the 
SOLEIL synchrotron \cite{ablett2019galaxies}. In brief, a vertically collimated 
x-ray beam (divergence $\sim$30$~\mu$rad, size $\sim$60$~\mu$m) illuminated the 
sample, which was positioned using a hexapod system capable of precise angular 
scanning. Reflected x-rays were collected using a 2D detector positioned in the 
forward direction, while fluorescence passing through a pinhole was recorded by 
a silicon drift detector (SDD) in the vertical direction. Cavity mode angles 
were identified using the reflectivity and TFY curves over the grazing angles 
\cite{supp2024}, and the TFY spectra were further used for validation of 
emission measurements \cite{supp2024}. We utilize eight VH analyzers to 
maximize the detection solid angle, with their emission detection zones aligned 
and overlapped across all analyzers \cite{supp2024}. Emission images were 
recorded by a dedicated 2D detector.

\begin{figure}[htbp]
	\centering
	\includegraphics[width=0.8\linewidth]{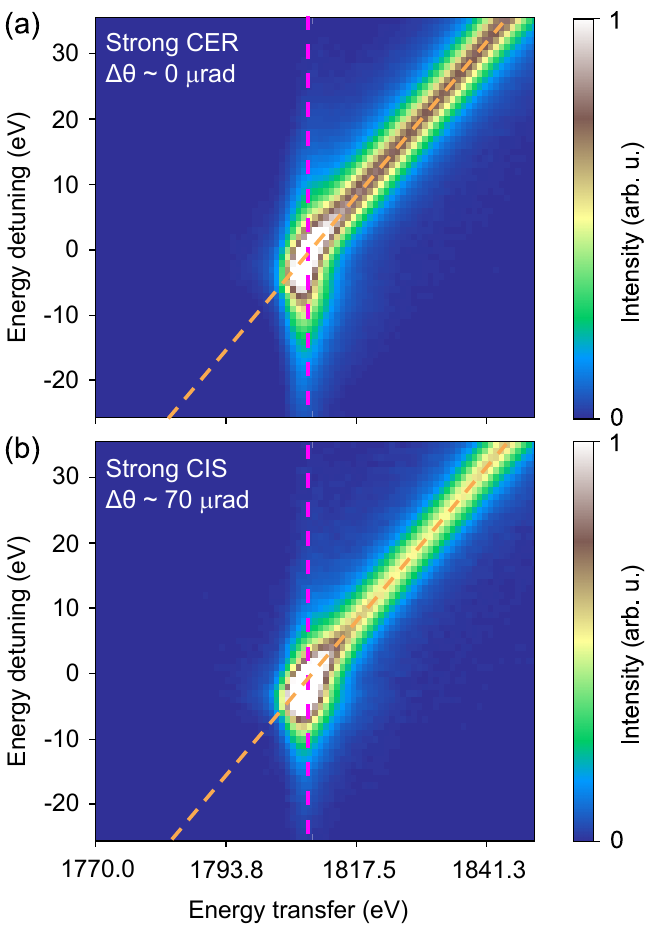}
	\caption{The RIXS planes displayed with energy transfer axis. (a) The 
	incident angle is at the first order of the cavity mode, where CER is 
	strong and CIS is weak. (b) The incident angle is set to have an offset of 
	$70~\mu$rad from (a), where CIS is strong and CER is weak. \rev{The 
	vertical pink dashed lines indicate the peak intensities of the Raman 
	feature.}
	\label{fig3} }
\end{figure}

\paragraph{Results and discussion}Figure \ref{fig3} presents the RIXS planes at 
the cavity mode ($\sim$3.5 mrad) and $70~\mu$rad offset angle, respectively. 
Compared to the RIXS plane recorded at a large angle (8.7 mrad, Fig. 
\ref{fig1}(c)), we observe a remarkable energy shift and line-shape broadening. 
Note that the resonant Raman peak (at constant energy transfer) at large angle, 
reflecting the natural decay width of the core-hole state, is centered at zero 
detuning energy in the absence of cavity effects. The constant emission energy 
feature (yellow dashed line), oriented along the 45\textdegree~diagonal, 
overlaps with the Raman signal (pink dashed line) near zero energy detuning, 
due to the emission linewidth and finite spectrometer resolution. Additionally, 
a slight energy offset is observed for the absorption edge jump, manifesting as 
a small extension towards the lower-energy transfer. This finding aligns with 
previous results, showing absorption edges shifted slightly above the white 
line transition determined from fittings to TFY spectra \cite{huang2021, 
ma2022, huang2024}.

At the cavity mode angle, the cavity effects strongly enhance the decay rate of 
the core-hole state, resulting in a visibly stretched profile feature, as 
evidenced by extended tails on both sides of zero detuning energy in Fig. 
\ref{fig3}(a). The profile broadening is more pronounced towards lower energy 
due to a finite negative cavity-induced energy shift ($\delta_c$). 
Additionally, the intensity of the constant emission energy feature increases 
significantly since the field inside the cavity is enhanced at the cavity mode, 
reflecting a standing wave effect \cite{zegenhagen2013}. Previous experiments, 
where only TFY spectra were measured \cite{haber2019, huang2021, huang2024}, 
suffered from severe overlap between resonant transitions and the absorption 
edge. Here, however, the overlap is notably reduced since the Raman feature 
broadens primarily along the vertical (energy-transfer) axis due to 
cavity-enhanced decay ($\gamma_c$). Although the cavity effects on the 
transitions to continuum states remain under debate \cite{huang2021, 
huang2024}, any such effect is expected to be weaker than for the resonant 
transitions. Our results clearly establish core-to-core RIXS as a powerful 
probe for x-ray quantum optics involving inner-shell transitions. Furthermore, 
the broadened Raman tails carry additional spectroscopic information, enabling 
novel applications. For example, at fixed incident x-ray energy, horizontal 
cuts across the RIXS plane yield spectra sensitive to final states emerging 
from a common intermediate core-hole state \cite{blachucki2014, baker2017}. The 
CER effect thus substantially improves the visibility of resonant transitions. 
Figure \ref{fig3}(b) shows the RIXS plane exhibiting a pronounced energy shift. 
Here, the Raman feature shifts to lower detuning energy and broadens slightly. 
Its overlap with the constant emission feature is further reduced since it 
moves further from the absorption edge. Such CIS can be exploited in HERFD 
x-ray absorption spectroscopy \cite{bauer2014}, where choosing a lower emission 
energy (negative $\delta_c$) enhances the visibility of bound-state features 
without sacrificing intensity (see \cite{supp2024} for details).

\begin{figure*}[htbp]
    \centering
    \includegraphics[width=1.0\linewidth]{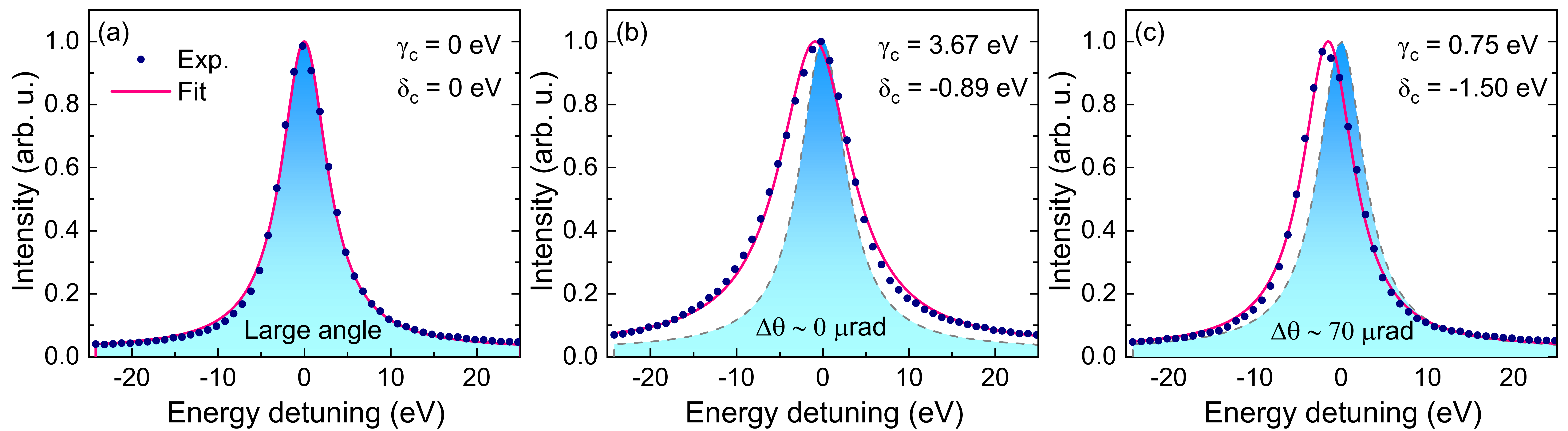}
    \caption{The extracted resonant spectrum of the Raman feature. The spectra 
    are integrated within a 2-eV band centered around the energy 
    transfer peak in the corresponding RIXS plane. (a) The incident angle is at 
    0.5\textdegree. (b) The incident angle is close to the cavity mode angle 
    ($\Delta\theta \sim 0~\mu$rad) and (c) detuned to an angle offset of about 
    $70~\mu$rad. The fittings are performed using a Lorentzian function with a 
    constant background. The shaded curves indicate the fitted profile of panel 
    (a). 
    \label{fig4}}
\end{figure*}

To explicitly investigate the cavity effects in the RIXS spectra, we extracted 
resonant transition profiles by integrating intensities within a \rev{$\pm 
1$}-eV window \rev{(2-eV band in total) centered} around the Raman peak 
\rev{(indicated by the pink dashed lines in Fig. \ref{fig3})}. The resulting 
spectra, fitted to Lorentzian profiles derived from Eq. (\ref{kheq}), are 
presented in Fig. \ref{fig4}. Such isolated resonances cannot be observed using 
conventional TFY or reflectivity measurements \cite{haber2019, huang2021, 
ma2022, huang2024}. Figure \ref{fig4}(a) provides a reference spectrum acquired 
at a large incident angle without cavity effects, from which we determine the 
cavity-induced $\delta_c$ and $\gamma_c$ as shown in Figs. \ref{fig4}(b) and 
(c). At the cavity mode angle, we observe a pronounced CER accompanied by a 
small CIS, consistent with theoretical predictions. Whereas we find a larger 
CIS and reduced CER at an offset angle of $\Delta \theta = 70~\mu$rad. The 
measured CER and CIS values are slightly smaller than theoretical predictions, 
likely due to residual angular divergence introduced by curvatures in the 
sample surface and the uncertainty in determining the exact cavity mode angle 
\cite{supp2024}. 

\rev{Additionally, we observe a subtle deviation from a symmetric Lorentzian 
profile at the cavity mode angle in Fig. \ref{fig4}(b). This asymmetry is not 
clearly visible at offset angles and is not captured by our present model. 
While various types of spectral asymmetries are known in quantum optics, such 
as Fano interference (e.g. \cite{ma2022} and references therein) and 
inhomogeneous atomic ensembles \cite{andreji2021}, the effect here may arise 
from the frequency-dependent response of the x-ray cavity itself 
\cite{huang2024}, which has not, to our knowledge, been reported in the context 
of inner-shell transitions. Further measurements across both positive and 
negative cavity detunings will be required to clarify the underlying mechanism.}


\paragraph{Summary}To summarize, we present the first measurement of core-to-core resonant inelastic scattering spectra from an x-ray cavity sample. Transitions to the resonant bound state and continuum state manifest as vertical and diagonal streaks in the RIXS planes, respectively. RIXS measurements were performed under a large incident angle (negligible cavity effects) and two cavity detunings chosen to introduce either a strong cavity-enhanced decay rate (CER) or a strong cavity-induced energy shift (CIS). We observe a pronounced spectral broadening of the Raman feature under strong CER conditions, while the Raman peak exhibits a clear shift towards lower energies under negative CIS conditions.

Our findings highlight the capability of core-to-core inelastic scattering in 
investigating x-ray cavity effects, particularly in distinguishing transitions 
to the resonant bound state and the continuum state. These results demonstrate 
that RIXS holds promise for exploring x-ray cavities with inner-shell 
transitions, drawing parallels to nuclear transitions in terms of collective 
effects \cite{rohlsberger2010, huang2024}, Rabi splitting \cite{haber2016, 
haber2017}, etc. Moreover, our results underscore the potential for 
experimental observables beyond elastic scattering - such as reflection and 
transmission - for cavity effect measurements \cite{heeg2013pra, heeg2015pra, 
kong2020, lentrodt2020prr, huang2024}, suggesting that observables like 
conversion electron detection and secondary fluorescence 
\cite{rohlsberger2004}, could offer comparable information and additional 
insights to nuclear transition systems \cite{röhlsberger2020, röhlsberger2021}. 
\rev{Extended applications can also be envisioned. For example, it has been 
pointed out that cavity-controlled RIXS may help suppress intermediate-state 
dynamics, which could, in turn, facilitate more quantitative theoretical 
modeling of RIXS.} Recent advances in diffraction-limited storage ring 
synchrotrons \cite{Einfeld2014}, seeded x-ray free-electron lasers 
\cite{liu2023} and attosecond x-ray pulses \cite{yan2024}, now provide 
unprecedented photon coherence, brilliance, and degeneracy, opening new 
avenues, such as the interplay between x-ray cavity controlled effects and 
elementary excitations with lower energy transfer \cite{ament2011} and 
nonlinear x-ray phenomena \cite{rohringer2019}.

\paragraph{Acknowledgement} The experiment was carried out with beam time approved by the GALAXIES beamline of SOLEIL synchrotron radiation facility (Proposal No. 20220927). The cavity sample was preliminarily tested with beam time at the B16 Test Beamline at the Diamond Light Source (Proposal No. MM31397). We acknowledge the support from Dr. Lingfei Hu (Diamond, now at USTC) during the B16 beamtime. We acknowledge the support of the National Natural Science Foundation of China through Grant No. 12334010, and R\&D HELIOS project at FXE of European XFEL. S.-X. W. acknowledges the support by the State of Hesse within the Research Cluster ELEMENTS (Project ID 500/10.006).

\bibliography{Refs.bib}

\begin{thebibliography}{11}%
\makeatletter
\providecommand \@ifxundefined [1]{%
 \@ifx{#1\undefined}
}%
\providecommand \@ifnum [1]{%
 \ifnum #1\expandafter \@firstoftwo
 \else \expandafter \@secondoftwo
 \fi
}%
\providecommand \@ifx [1]{%
 \ifx #1\expandafter \@firstoftwo
 \else \expandafter \@secondoftwo
 \fi
}%
\providecommand \natexlab [1]{#1}%
\providecommand \enquote  [1]{``#1''}%
\providecommand \bibnamefont  [1]{#1}%
\providecommand \bibfnamefont [1]{#1}%
\providecommand \citenamefont [1]{#1}%
\providecommand \href@noop [0]{\@secondoftwo}%
\providecommand \href [0]{\begingroup \@sanitize@url \@href}%
\providecommand \@href[1]{\@@startlink{#1}\@@href}%
\providecommand \@@href[1]{\endgroup#1\@@endlink}%
\providecommand \@sanitize@url [0]{\catcode `\\12\catcode `\$12\catcode
  `\&12\catcode `\#12\catcode `\^12\catcode `\_12\catcode `\%12\relax}%
\providecommand \@@startlink[1]{}%
\providecommand \@@endlink[0]{}%
\providecommand \url  [0]{\begingroup\@sanitize@url \@url }%
\providecommand \@url [1]{\endgroup\@href {#1}{\urlprefix }}%
\providecommand \urlprefix  [0]{URL }%
\providecommand \Eprint [0]{\href }%
\providecommand \doibase [0]{https://doi.org/}%
\providecommand \selectlanguage [0]{\@gobble}%
\providecommand \bibinfo  [0]{\@secondoftwo}%
\providecommand \bibfield  [0]{\@secondoftwo}%
\providecommand \translation [1]{[#1]}%
\providecommand \BibitemOpen [0]{}%
\providecommand \bibitemStop [0]{}%
\providecommand \bibitemNoStop [0]{.\EOS\space}%
\providecommand \EOS [0]{\spacefactor3000\relax}%
\providecommand \BibitemShut  [1]{\csname bibitem#1\endcsname}%
\let\auto@bib@innerbib\@empty
\bibitem [{\citenamefont {Kotani}\ and\ \citenamefont
  {Shin}(2001)}]{kotani2001}%
  \BibitemOpen
  \bibfield  {author} {\bibinfo {author} {\bibfnamefont {A.}~\bibnamefont
  {Kotani}}\ and\ \bibinfo {author} {\bibfnamefont {S.}~\bibnamefont {Shin}},\
  }\href {https://doi.org/10.1103/RevModPhys.73.203} {\bibfield  {journal}
  {\bibinfo  {journal} {Rev. Mod. Phys.}\ }\textbf {\bibinfo {volume} {73}},\
  \bibinfo {pages} {203} (\bibinfo {year} {2001})}\BibitemShut {NoStop}%
\bibitem [{\citenamefont {Ament}\ \emph {et~al.}(2011)\citenamefont {Ament},
  \citenamefont {van Veenendaal}, \citenamefont {Devereaux}, \citenamefont
  {Hill},\ and\ \citenamefont {van~den Brink}}]{ament2011}%
  \BibitemOpen
  \bibfield  {author} {\bibinfo {author} {\bibfnamefont {L.~J.~P.}\
  \bibnamefont {Ament}}, \bibinfo {author} {\bibfnamefont {M.}~\bibnamefont
  {van Veenendaal}}, \bibinfo {author} {\bibfnamefont {T.~P.}\ \bibnamefont
  {Devereaux}}, \bibinfo {author} {\bibfnamefont {J.~P.}\ \bibnamefont
  {Hill}},\ and\ \bibinfo {author} {\bibfnamefont {J.}~\bibnamefont {van~den
  Brink}},\ }\href {https://doi.org/10.1103/RevModPhys.83.705} {\bibfield
  {journal} {\bibinfo  {journal} {Rev. Mod. Phys.}\ }\textbf {\bibinfo {volume}
  {83}},\ \bibinfo {pages} {705} (\bibinfo {year} {2011})}\BibitemShut
  {NoStop}%
\bibitem [{\citenamefont {R\"ohlsberger}\ \emph {et~al.}(2004)\citenamefont
  {R\"ohlsberger}, \citenamefont {Klein}, \citenamefont {Schlage},
  \citenamefont {Leupold},\ and\ \citenamefont
  {R\"uffer}}]{rohlsberger2004prb}%
  \BibitemOpen
  \bibfield  {author} {\bibinfo {author} {\bibfnamefont {R.}~\bibnamefont
  {R\"ohlsberger}}, \bibinfo {author} {\bibfnamefont {T.}~\bibnamefont
  {Klein}}, \bibinfo {author} {\bibfnamefont {K.}~\bibnamefont {Schlage}},
  \bibinfo {author} {\bibfnamefont {O.}~\bibnamefont {Leupold}},\ and\ \bibinfo
  {author} {\bibfnamefont {R.}~\bibnamefont {R\"uffer}},\ }\href
  {https://doi.org/10.1103/PhysRevB.69.235412} {\bibfield  {journal} {\bibinfo
  {journal} {Phys. Rev. B}\ }\textbf {\bibinfo {volume} {69}},\ \bibinfo
  {pages} {235412} (\bibinfo {year} {2004})}\BibitemShut {NoStop}%
\bibitem [{\citenamefont {V.~H{\'a}mos}(1933)}]{hamos1933}%
  \BibitemOpen
  \bibfield  {author} {\bibinfo {author} {\bibfnamefont {L.}~\bibnamefont
  {V.~H{\'a}mos}},\ }\href@noop {} {\bibfield  {journal} {\bibinfo  {journal}
  {Ann. Phys. (Leipzig)}\ }\textbf {\bibinfo {volume} {409}},\ \bibinfo {pages}
  {716} (\bibinfo {year} {1933})}\BibitemShut {NoStop}%
\bibitem [{\citenamefont {Wach}\ \emph {et~al.}(2020)\citenamefont {Wach},
  \citenamefont {S{\'{a}}},\ and\ \citenamefont {Szlachetko}}]{wach2020}%
  \BibitemOpen
  \bibfield  {author} {\bibinfo {author} {\bibfnamefont {A.}~\bibnamefont
  {Wach}}, \bibinfo {author} {\bibfnamefont {J.}~\bibnamefont {S{\'{a}}}},\
  and\ \bibinfo {author} {\bibfnamefont {J.}~\bibnamefont {Szlachetko}},\
  }\href {https://doi.org/10.1107/S1600577520003690} {\bibfield  {journal}
  {\bibinfo  {journal} {J. Synchrotron Rad.}\ }\textbf {\bibinfo {volume}
  {27}},\ \bibinfo {pages} {689} (\bibinfo {year} {2020})}\BibitemShut
  {NoStop}%
\bibitem [{\citenamefont {Zhao}\ \emph {et~al.}(2025)\citenamefont {Zhao},
  \citenamefont {Wang}, \citenamefont {Wang}, \citenamefont {Su}, \citenamefont
  {Ma}, \citenamefont {Huang},\ and\ \citenamefont {Zhu}}]{zhao2025}%
  \BibitemOpen
  \bibfield  {author} {\bibinfo {author} {\bibfnamefont {Z.~Q.}\ \bibnamefont
  {Zhao}}, \bibinfo {author} {\bibfnamefont {S.~X.}\ \bibnamefont {Wang}},
  \bibinfo {author} {\bibfnamefont {X.~Y.}\ \bibnamefont {Wang}}, \bibinfo
  {author} {\bibfnamefont {Y.}~\bibnamefont {Su}}, \bibinfo {author}
  {\bibfnamefont {Z.~R.}\ \bibnamefont {Ma}}, \bibinfo {author} {\bibfnamefont
  {X.~C.}\ \bibnamefont {Huang}},\ and\ \bibinfo {author} {\bibfnamefont
  {L.~F.}\ \bibnamefont {Zhu}},\ }\href
  {https://doi.org/10.7498/aps.74.20250659} {\bibfield  {journal} {\bibinfo
  {journal} {Acta Phys. Sin.}\ }\textbf {\bibinfo {volume} {74}},\ \bibinfo
  {pages} {183201} (\bibinfo {year} {2025})}\BibitemShut {NoStop}%
\bibitem [{\citenamefont {Ma}\ \emph {et~al.}(2022)\citenamefont {Ma},
  \citenamefont {Huang}, \citenamefont {Li}, \citenamefont {Wang},
  \citenamefont {Liu}, \citenamefont {Wang}, \citenamefont {Li}, \citenamefont
  {Li},\ and\ \citenamefont {Zhu}}]{ma2022}%
  \BibitemOpen
  \bibfield  {author} {\bibinfo {author} {\bibfnamefont {Z.-R.}\ \bibnamefont
  {Ma}}, \bibinfo {author} {\bibfnamefont {X.-C.}\ \bibnamefont {Huang}},
  \bibinfo {author} {\bibfnamefont {T.-J.}\ \bibnamefont {Li}}, \bibinfo
  {author} {\bibfnamefont {H.-C.}\ \bibnamefont {Wang}}, \bibinfo {author}
  {\bibfnamefont {G.-C.}\ \bibnamefont {Liu}}, \bibinfo {author} {\bibfnamefont
  {Z.-S.}\ \bibnamefont {Wang}}, \bibinfo {author} {\bibfnamefont
  {B.}~\bibnamefont {Li}}, \bibinfo {author} {\bibfnamefont {W.-B.}\
  \bibnamefont {Li}},\ and\ \bibinfo {author} {\bibfnamefont {L.-F.}\
  \bibnamefont {Zhu}},\ }\href {https://doi.org/10.1103/PhysRevLett.129.213602}
  {\bibfield  {journal} {\bibinfo  {journal} {Phys. Rev. Lett.}\ }\textbf
  {\bibinfo {volume} {129}},\ \bibinfo {pages} {213602} (\bibinfo {year}
  {2022})}\BibitemShut {NoStop}%
\bibitem [{\citenamefont {Bj{\"{o}}rck}\ and\ \citenamefont
  {Andersson}(2007)}]{genx2007}%
  \BibitemOpen
  \bibfield  {author} {\bibinfo {author} {\bibfnamefont {M.}~\bibnamefont
  {Bj{\"{o}}rck}}\ and\ \bibinfo {author} {\bibfnamefont {G.}~\bibnamefont
  {Andersson}},\ }\href {https://doi.org/10.1107/S0021889807045086} {\bibfield
  {journal} {\bibinfo  {journal} {J. Appl. Crystallogr.}\ }\textbf {\bibinfo
  {volume} {40}},\ \bibinfo {pages} {1174} (\bibinfo {year}
  {2007})}\BibitemShut {NoStop}%
\bibitem [{\citenamefont {B\l{}achucki}\ \emph {et~al.}(2014)\citenamefont
  {B\l{}achucki}, \citenamefont {Szlachetko}, \citenamefont {Hoszowska},
  \citenamefont {Dousse}, \citenamefont {Kayser}, \citenamefont {Nachtegaal},\
  and\ \citenamefont {S\'a}}]{blachucki2014}%
  \BibitemOpen
  \bibfield  {author} {\bibinfo {author} {\bibfnamefont {W.}~\bibnamefont
  {B\l{}achucki}}, \bibinfo {author} {\bibfnamefont {J.}~\bibnamefont
  {Szlachetko}}, \bibinfo {author} {\bibfnamefont {J.}~\bibnamefont
  {Hoszowska}}, \bibinfo {author} {\bibfnamefont {J.-C.}\ \bibnamefont
  {Dousse}}, \bibinfo {author} {\bibfnamefont {Y.}~\bibnamefont {Kayser}},
  \bibinfo {author} {\bibfnamefont {M.}~\bibnamefont {Nachtegaal}},\ and\
  \bibinfo {author} {\bibfnamefont {J.}~\bibnamefont {S\'a}},\ }\href
  {https://doi.org/10.1103/PhysRevLett.112.173003} {\bibfield  {journal}
  {\bibinfo  {journal} {Phys. Rev. Lett.}\ }\textbf {\bibinfo {volume} {112}},\
  \bibinfo {pages} {173003} (\bibinfo {year} {2014})}\BibitemShut {NoStop}%
\bibitem [{\citenamefont {Bauer}(2014)}]{bauer2014}%
  \BibitemOpen
  \bibfield  {author} {\bibinfo {author} {\bibfnamefont {M.}~\bibnamefont
  {Bauer}},\ }\href {https://doi.org/10.1039/C4CP00904E} {\bibfield  {journal}
  {\bibinfo  {journal} {Phys. Chem. Chem. Phys.}\ }\textbf {\bibinfo {volume}
  {16}},\ \bibinfo {pages} {13827} (\bibinfo {year} {2014})}\BibitemShut
  {NoStop}%
\bibitem [{\citenamefont {H\"am\"al\"ainen}\ \emph {et~al.}(1991)\citenamefont
  {H\"am\"al\"ainen}, \citenamefont {Siddons}, \citenamefont {Hastings},\ and\
  \citenamefont {Berman}}]{Hämäläinen1991}%
  \BibitemOpen
  \bibfield  {author} {\bibinfo {author} {\bibfnamefont {K.}~\bibnamefont
  {H\"am\"al\"ainen}}, \bibinfo {author} {\bibfnamefont {D.~P.}\ \bibnamefont
  {Siddons}}, \bibinfo {author} {\bibfnamefont {J.~B.}\ \bibnamefont
  {Hastings}},\ and\ \bibinfo {author} {\bibfnamefont {L.~E.}\ \bibnamefont
  {Berman}},\ }\href {https://doi.org/10.1103/PhysRevLett.67.2850} {\bibfield
  {journal} {\bibinfo  {journal} {Phys. Rev. Lett.}\ }\textbf {\bibinfo
  {volume} {67}},\ \bibinfo {pages} {2850} (\bibinfo {year}
  {1991})}\BibitemShut {NoStop}%
\end{thebibliography}%


\begin{thebibliography}{57}%
\makeatletter
\providecommand \@ifxundefined [1]{%
 \@ifx{#1\undefined}
}%
\providecommand \@ifnum [1]{%
 \ifnum #1\expandafter \@firstoftwo
 \else \expandafter \@secondoftwo
 \fi
}%
\providecommand \@ifx [1]{%
 \ifx #1\expandafter \@firstoftwo
 \else \expandafter \@secondoftwo
 \fi
}%
\providecommand \natexlab [1]{#1}%
\providecommand \enquote  [1]{``#1''}%
\providecommand \bibnamefont  [1]{#1}%
\providecommand \bibfnamefont [1]{#1}%
\providecommand \citenamefont [1]{#1}%
\providecommand \href@noop [0]{\@secondoftwo}%
\providecommand \href [0]{\begingroup \@sanitize@url \@href}%
\providecommand \@href[1]{\@@startlink{#1}\@@href}%
\providecommand \@@href[1]{\endgroup#1\@@endlink}%
\providecommand \@sanitize@url [0]{\catcode `\\12\catcode `\$12\catcode
  `\&12\catcode `\#12\catcode `\^12\catcode `\_12\catcode `\%12\relax}%
\providecommand \@@startlink[1]{}%
\providecommand \@@endlink[0]{}%
\providecommand \url  [0]{\begingroup\@sanitize@url \@url }%
\providecommand \@url [1]{\endgroup\@href {#1}{\urlprefix }}%
\providecommand \urlprefix  [0]{URL }%
\providecommand \Eprint [0]{\href }%
\providecommand \doibase [0]{https://doi.org/}%
\providecommand \selectlanguage [0]{\@gobble}%
\providecommand \bibinfo  [0]{\@secondoftwo}%
\providecommand \bibfield  [0]{\@secondoftwo}%
\providecommand \translation [1]{[#1]}%
\providecommand \BibitemOpen [0]{}%
\providecommand \bibitemStop [0]{}%
\providecommand \bibitemNoStop [0]{.\EOS\space}%
\providecommand \EOS [0]{\spacefactor3000\relax}%
\providecommand \BibitemShut  [1]{\csname bibitem#1\endcsname}%
\let\auto@bib@innerbib\@empty
\bibitem [{\citenamefont {Adams}\ \emph {et~al.}(2013)\citenamefont {Adams},
  \citenamefont {Buth}, \citenamefont {Cavaletto} \emph {et~al.}}]{adams2013}%
  \BibitemOpen
  \bibfield  {author} {\bibinfo {author} {\bibfnamefont {B.~W.}\ \bibnamefont
  {Adams}}, \bibinfo {author} {\bibfnamefont {C.}~\bibnamefont {Buth}},
  \bibinfo {author} {\bibfnamefont {S.~M.}\ \bibnamefont {Cavaletto}}, \emph
  {et~al.},\ }\href {https://doi.org/10.1080/09500340.2012.752113} {\bibfield
  {journal} {\bibinfo  {journal} {J. Mod. Optic.}\ }\textbf {\bibinfo {volume}
  {60}},\ \bibinfo {pages} {2} (\bibinfo {year} {2013})}\BibitemShut {NoStop}%
\bibitem [{\citenamefont {Kuznetsova}\ and\ \citenamefont
  {Kocharovskaya}(2017)}]{kuznetsova2017}%
  \BibitemOpen
  \bibfield  {author} {\bibinfo {author} {\bibfnamefont {E.}~\bibnamefont
  {Kuznetsova}}\ and\ \bibinfo {author} {\bibfnamefont {O.}~\bibnamefont
  {Kocharovskaya}},\ }\href {https://doi.org/10.1038/s41566-017-0034-y}
  {\bibfield  {journal} {\bibinfo  {journal} {Nat. Photon.}\ }\textbf {\bibinfo
  {volume} {11}},\ \bibinfo {pages} {685} (\bibinfo {year} {2017})}\BibitemShut
  {NoStop}%
\bibitem [{\citenamefont {Wong}\ and\ \citenamefont
  {Kaminer}(2021)}]{wong2021}%
  \BibitemOpen
  \bibfield  {author} {\bibinfo {author} {\bibfnamefont {L.~J.}\ \bibnamefont
  {Wong}}\ and\ \bibinfo {author} {\bibfnamefont {I.}~\bibnamefont {Kaminer}},\
  }\href {https://doi.org/10.1063/5.0060552} {\bibfield  {journal} {\bibinfo
  {journal} {Appl. Phys. Lett.}\ }\textbf {\bibinfo {volume} {119}},\ \bibinfo
  {pages} {130502} (\bibinfo {year} {2021})}\BibitemShut {NoStop}%
\bibitem [{\citenamefont {Wang}\ \emph {et~al.}(2024)\citenamefont {Wang},
  \citenamefont {Li}, \citenamefont {Huang},\ and\ \citenamefont
  {Zhu}}]{wang2024acta}%
  \BibitemOpen
  \bibfield  {author} {\bibinfo {author} {\bibfnamefont {S.-X.}\ \bibnamefont
  {Wang}}, \bibinfo {author} {\bibfnamefont {T.-J.}\ \bibnamefont {Li}},
  \bibinfo {author} {\bibfnamefont {X.-C.}\ \bibnamefont {Huang}},\ and\
  \bibinfo {author} {\bibfnamefont {L.-F.}\ \bibnamefont {Zhu}},\ }\href
  {https://doi.org/10.7498/aps.73.20241218} {\bibfield  {journal} {\bibinfo
  {journal} {Acta Phys. Sin.}\ }\textbf {\bibinfo {volume} {73}},\ \bibinfo
  {pages} {246101} (\bibinfo {year} {2024})}\BibitemShut {NoStop}%
\bibitem [{\citenamefont {Bostedt}\ \emph {et~al.}(2016)\citenamefont
  {Bostedt}, \citenamefont {Boutet}, \citenamefont {Fritz}, \citenamefont
  {Huang}, \citenamefont {Lee}, \citenamefont {Lemke}, \citenamefont {Robert},
  \citenamefont {Schlotter}, \citenamefont {Turner},\ and\ \citenamefont
  {Williams}}]{Bostedt2011}%
  \BibitemOpen
  \bibfield  {author} {\bibinfo {author} {\bibfnamefont {C.}~\bibnamefont
  {Bostedt}}, \bibinfo {author} {\bibfnamefont {S.}~\bibnamefont {Boutet}},
  \bibinfo {author} {\bibfnamefont {D.~M.}\ \bibnamefont {Fritz}}, \bibinfo
  {author} {\bibfnamefont {Z.}~\bibnamefont {Huang}}, \bibinfo {author}
  {\bibfnamefont {H.~J.}\ \bibnamefont {Lee}}, \bibinfo {author} {\bibfnamefont
  {H.~T.}\ \bibnamefont {Lemke}}, \bibinfo {author} {\bibfnamefont
  {A.}~\bibnamefont {Robert}}, \bibinfo {author} {\bibfnamefont {W.~F.}\
  \bibnamefont {Schlotter}}, \bibinfo {author} {\bibfnamefont {J.~J.}\
  \bibnamefont {Turner}},\ and\ \bibinfo {author} {\bibfnamefont {G.~J.}\
  \bibnamefont {Williams}},\ }\href
  {https://doi.org/10.1103/RevModPhys.88.015007} {\bibfield  {journal}
  {\bibinfo  {journal} {Rev. Mod. Phys.}\ }\textbf {\bibinfo {volume} {88}},\
  \bibinfo {pages} {015007} (\bibinfo {year} {2016})}\BibitemShut {NoStop}%
\bibitem [{\citenamefont {Einfeld}\ \emph {et~al.}(2014)\citenamefont
  {Einfeld}, \citenamefont {Plesko},\ and\ \citenamefont
  {Schaper}}]{Einfeld2014}%
  \BibitemOpen
  \bibfield  {author} {\bibinfo {author} {\bibfnamefont {D.}~\bibnamefont
  {Einfeld}}, \bibinfo {author} {\bibfnamefont {M.}~\bibnamefont {Plesko}},\
  and\ \bibinfo {author} {\bibfnamefont {J.}~\bibnamefont {Schaper}},\ }\href
  {https://doi.org/10.1107/S160057751401193X} {\bibfield  {journal} {\bibinfo
  {journal} {J. Synchrotron Rad.}\ }\textbf {\bibinfo {volume} {21}},\ \bibinfo
  {pages} {856} (\bibinfo {year} {2014})}\BibitemShut {NoStop}%
\bibitem [{\citenamefont {R{\"o}hlsberger}\ \emph {et~al.}(2020)\citenamefont
  {R{\"o}hlsberger}, \citenamefont {Evers},\ and\ \citenamefont
  {Shwartz}}]{röhlsberger2020}%
  \BibitemOpen
  \bibfield  {author} {\bibinfo {author} {\bibfnamefont {R.}~\bibnamefont
  {R{\"o}hlsberger}}, \bibinfo {author} {\bibfnamefont {J.}~\bibnamefont
  {Evers}},\ and\ \bibinfo {author} {\bibfnamefont {S.}~\bibnamefont
  {Shwartz}},\ }\bibinfo {title} {Quantum and nonlinear optics with hard
  x-rays},\ in\ \href {https://doi.org/10.1007/978-3-030-23201-6_32} {\emph
  {\bibinfo {booktitle} {Synchrotron Light Sources and Free-Electron Lasers:
  Accelerator Physics, Instrumentation and Science Applications}}},\ \bibinfo
  {editor} {edited by\ \bibinfo {editor} {\bibfnamefont {E.~J.}\ \bibnamefont
  {Jaeschke}}, \bibinfo {editor} {\bibfnamefont {S.}~\bibnamefont {Khan}},
  \bibinfo {editor} {\bibfnamefont {J.~R.}\ \bibnamefont {Schneider}},\ and\
  \bibinfo {editor} {\bibfnamefont {J.~B.}\ \bibnamefont {Hastings}}}\
  (\bibinfo  {publisher} {Springer International Publishing},\ \bibinfo
  {address} {Cham},\ \bibinfo {year} {2020})\ pp.\ \bibinfo {pages}
  {1399--1431}\BibitemShut {NoStop}%
\bibitem [{\citenamefont {R{\"o}hlsberger}\ and\ \citenamefont
  {Evers}(2021)}]{röhlsberger2021}%
  \BibitemOpen
  \bibfield  {author} {\bibinfo {author} {\bibfnamefont {R.}~\bibnamefont
  {R{\"o}hlsberger}}\ and\ \bibinfo {author} {\bibfnamefont {J.}~\bibnamefont
  {Evers}},\ }\href@noop {} {\bibfield  {journal} {\bibinfo  {journal} {Modern
  M{\"o}ssbauer Spectroscopy}\ ,\ \bibinfo {pages} {105}} (\bibinfo {year}
  {2021})}\BibitemShut {NoStop}%
\bibitem [{\citenamefont {Raimond}\ \emph {et~al.}(2001)\citenamefont
  {Raimond}, \citenamefont {Brune},\ and\ \citenamefont
  {Haroche}}]{raimond2001}%
  \BibitemOpen
  \bibfield  {author} {\bibinfo {author} {\bibfnamefont {J.~M.}\ \bibnamefont
  {Raimond}}, \bibinfo {author} {\bibfnamefont {M.}~\bibnamefont {Brune}},\
  and\ \bibinfo {author} {\bibfnamefont {S.}~\bibnamefont {Haroche}},\ }\href
  {https://doi.org/10.1103/RevModPhys.73.565} {\bibfield  {journal} {\bibinfo
  {journal} {Rev. Mod. Phys.}\ }\textbf {\bibinfo {volume} {73}},\ \bibinfo
  {pages} {565} (\bibinfo {year} {2001})}\BibitemShut {NoStop}%
\bibitem [{\citenamefont {R\"ohlsberger}\ \emph {et~al.}(2005)\citenamefont
  {R\"ohlsberger}, \citenamefont {Schlage}, \citenamefont {Klein},\ and\
  \citenamefont {Leupold}}]{rohlsberger2005}%
  \BibitemOpen
  \bibfield  {author} {\bibinfo {author} {\bibfnamefont {R.}~\bibnamefont
  {R\"ohlsberger}}, \bibinfo {author} {\bibfnamefont {K.}~\bibnamefont
  {Schlage}}, \bibinfo {author} {\bibfnamefont {T.}~\bibnamefont {Klein}},\
  and\ \bibinfo {author} {\bibfnamefont {O.}~\bibnamefont {Leupold}},\ }\href
  {https://doi.org/10.1103/PhysRevLett.95.097601} {\bibfield  {journal}
  {\bibinfo  {journal} {Phys. Rev. Lett.}\ }\textbf {\bibinfo {volume} {95}},\
  \bibinfo {pages} {097601} (\bibinfo {year} {2005})}\BibitemShut {NoStop}%
\bibitem [{\citenamefont {Röhlsberger}\ \emph {et~al.}(2010)\citenamefont
  {Röhlsberger}, \citenamefont {Schlage}, \citenamefont {Sahoo}, \citenamefont
  {Couet},\ and\ \citenamefont {Rüffer}}]{rohlsberger2010}%
  \BibitemOpen
  \bibfield  {author} {\bibinfo {author} {\bibfnamefont {R.}~\bibnamefont
  {Röhlsberger}}, \bibinfo {author} {\bibfnamefont {K.}~\bibnamefont
  {Schlage}}, \bibinfo {author} {\bibfnamefont {B.}~\bibnamefont {Sahoo}},
  \bibinfo {author} {\bibfnamefont {S.}~\bibnamefont {Couet}},\ and\ \bibinfo
  {author} {\bibfnamefont {R.}~\bibnamefont {Rüffer}},\ }\href
  {https://doi.org/10.1126/science.1187770} {\bibfield  {journal} {\bibinfo
  {journal} {Science}\ }\textbf {\bibinfo {volume} {328}},\ \bibinfo {pages}
  {1248} (\bibinfo {year} {2010})}\BibitemShut {NoStop}%
\bibitem [{\citenamefont {R{\"o}hlsberger}\ \emph {et~al.}(2012)\citenamefont
  {R{\"o}hlsberger}, \citenamefont {Wille}, \citenamefont {Schlage},\ and\
  \citenamefont {Sahoo}}]{rohlsberger2012}%
  \BibitemOpen
  \bibfield  {author} {\bibinfo {author} {\bibfnamefont {R.}~\bibnamefont
  {R{\"o}hlsberger}}, \bibinfo {author} {\bibfnamefont {H.-C.}\ \bibnamefont
  {Wille}}, \bibinfo {author} {\bibfnamefont {K.}~\bibnamefont {Schlage}},\
  and\ \bibinfo {author} {\bibfnamefont {B.}~\bibnamefont {Sahoo}},\ }\href
  {https://doi.org/10.1038/nature10741} {\bibfield  {journal} {\bibinfo
  {journal} {Nature}\ }\textbf {\bibinfo {volume} {482}},\ \bibinfo {pages}
  {199} (\bibinfo {year} {2012})}\BibitemShut {NoStop}%
\bibitem [{\citenamefont {Heeg}\ \emph
  {et~al.}(2015{\natexlab{a}})\citenamefont {Heeg}, \citenamefont {Ott},
  \citenamefont {Schumacher}, \citenamefont {Wille}, \citenamefont
  {R\"ohlsberger}, \citenamefont {Pfeifer},\ and\ \citenamefont
  {Evers}}]{heeg2015fano}%
  \BibitemOpen
  \bibfield  {author} {\bibinfo {author} {\bibfnamefont {K.~P.}\ \bibnamefont
  {Heeg}}, \bibinfo {author} {\bibfnamefont {C.}~\bibnamefont {Ott}}, \bibinfo
  {author} {\bibfnamefont {D.}~\bibnamefont {Schumacher}}, \bibinfo {author}
  {\bibfnamefont {H.-C.}\ \bibnamefont {Wille}}, \bibinfo {author}
  {\bibfnamefont {R.}~\bibnamefont {R\"ohlsberger}}, \bibinfo {author}
  {\bibfnamefont {T.}~\bibnamefont {Pfeifer}},\ and\ \bibinfo {author}
  {\bibfnamefont {J.}~\bibnamefont {Evers}},\ }\href
  {https://doi.org/10.1103/PhysRevLett.114.207401} {\bibfield  {journal}
  {\bibinfo  {journal} {Phys. Rev. Lett.}\ }\textbf {\bibinfo {volume} {114}},\
  \bibinfo {pages} {207401} (\bibinfo {year} {2015}{\natexlab{a}})}\BibitemShut
  {NoStop}%
\bibitem [{\citenamefont {Heeg}\ \emph
  {et~al.}(2015{\natexlab{b}})\citenamefont {Heeg}, \citenamefont {Haber},
  \citenamefont {Schumacher}, \citenamefont {Bocklage}, \citenamefont {Wille},
  \citenamefont {Schulze}, \citenamefont {Loetzsch}, \citenamefont {Uschmann},
  \citenamefont {Paulus}, \citenamefont {R\"uffer}, \citenamefont
  {R\"ohlsberger},\ and\ \citenamefont {Evers}}]{heeg2015slow}%
  \BibitemOpen
  \bibfield  {author} {\bibinfo {author} {\bibfnamefont {K.~P.}\ \bibnamefont
  {Heeg}}, \bibinfo {author} {\bibfnamefont {J.}~\bibnamefont {Haber}},
  \bibinfo {author} {\bibfnamefont {D.}~\bibnamefont {Schumacher}}, \bibinfo
  {author} {\bibfnamefont {L.}~\bibnamefont {Bocklage}}, \bibinfo {author}
  {\bibfnamefont {H.-C.}\ \bibnamefont {Wille}}, \bibinfo {author}
  {\bibfnamefont {K.~S.}\ \bibnamefont {Schulze}}, \bibinfo {author}
  {\bibfnamefont {R.}~\bibnamefont {Loetzsch}}, \bibinfo {author}
  {\bibfnamefont {I.}~\bibnamefont {Uschmann}}, \bibinfo {author}
  {\bibfnamefont {G.~G.}\ \bibnamefont {Paulus}}, \bibinfo {author}
  {\bibfnamefont {R.}~\bibnamefont {R\"uffer}}, \bibinfo {author}
  {\bibfnamefont {R.}~\bibnamefont {R\"ohlsberger}},\ and\ \bibinfo {author}
  {\bibfnamefont {J.}~\bibnamefont {Evers}},\ }\href
  {https://doi.org/10.1103/PhysRevLett.114.203601} {\bibfield  {journal}
  {\bibinfo  {journal} {Phys. Rev. Lett.}\ }\textbf {\bibinfo {volume} {114}},\
  \bibinfo {pages} {203601} (\bibinfo {year} {2015}{\natexlab{b}})}\BibitemShut
  {NoStop}%
\bibitem [{\citenamefont {Haber}\ \emph {et~al.}(2017)\citenamefont {Haber},
  \citenamefont {Kong}, \citenamefont {Strohm}, \citenamefont {Willing},
  \citenamefont {Gollwitzer}, \citenamefont {Bocklage}, \citenamefont
  {Rüffer}, \citenamefont {Pálffy},\ and\ \citenamefont
  {Röhlsberger}}]{haber2017}%
  \BibitemOpen
  \bibfield  {author} {\bibinfo {author} {\bibfnamefont {J.}~\bibnamefont
  {Haber}}, \bibinfo {author} {\bibfnamefont {X.}~\bibnamefont {Kong}},
  \bibinfo {author} {\bibfnamefont {C.}~\bibnamefont {Strohm}}, \bibinfo
  {author} {\bibfnamefont {S.}~\bibnamefont {Willing}}, \bibinfo {author}
  {\bibfnamefont {J.}~\bibnamefont {Gollwitzer}}, \bibinfo {author}
  {\bibfnamefont {L.}~\bibnamefont {Bocklage}}, \bibinfo {author}
  {\bibfnamefont {R.}~\bibnamefont {Rüffer}}, \bibinfo {author} {\bibfnamefont
  {A.}~\bibnamefont {Pálffy}},\ and\ \bibinfo {author} {\bibfnamefont
  {R.}~\bibnamefont {Röhlsberger}},\ }\href
  {https://doi.org/10.1038/s41566-017-0013-3} {\bibfield  {journal} {\bibinfo
  {journal} {Nat. Photon.}\ }\textbf {\bibinfo {volume} {11}},\ \bibinfo
  {pages} {720} (\bibinfo {year} {2017})}\BibitemShut {NoStop}%
\bibitem [{\citenamefont {Huang}\ \emph {et~al.}(2017)\citenamefont {Huang},
  \citenamefont {Li}, \citenamefont {Kong},\ and\ \citenamefont
  {Zhu}}]{huang2017}%
  \BibitemOpen
  \bibfield  {author} {\bibinfo {author} {\bibfnamefont {X.-C.}\ \bibnamefont
  {Huang}}, \bibinfo {author} {\bibfnamefont {W.-B.}\ \bibnamefont {Li}},
  \bibinfo {author} {\bibfnamefont {X.-J.}\ \bibnamefont {Kong}},\ and\
  \bibinfo {author} {\bibfnamefont {L.-F.}\ \bibnamefont {Zhu}},\ }\href
  {https://doi.org/10.1364/OE.25.031337} {\bibfield  {journal} {\bibinfo
  {journal} {Opt. Express}\ }\textbf {\bibinfo {volume} {25}},\ \bibinfo
  {pages} {31337} (\bibinfo {year} {2017})}\BibitemShut {NoStop}%
\bibitem [{\citenamefont {Velten}\ \emph {et~al.}(2024)\citenamefont {Velten},
  \citenamefont {Bocklage}, \citenamefont {Zhang}, \citenamefont {Schlage},
  \citenamefont {Panchwanee}, \citenamefont {Sadashivaiah}, \citenamefont
  {Sergeev}, \citenamefont {Leupold}, \citenamefont {Chumakov}, \citenamefont
  {Kocharovskaya} \emph {et~al.}}]{velten2024}%
  \BibitemOpen
  \bibfield  {author} {\bibinfo {author} {\bibfnamefont {S.}~\bibnamefont
  {Velten}}, \bibinfo {author} {\bibfnamefont {L.}~\bibnamefont {Bocklage}},
  \bibinfo {author} {\bibfnamefont {X.}~\bibnamefont {Zhang}}, \bibinfo
  {author} {\bibfnamefont {K.}~\bibnamefont {Schlage}}, \bibinfo {author}
  {\bibfnamefont {A.}~\bibnamefont {Panchwanee}}, \bibinfo {author}
  {\bibfnamefont {S.}~\bibnamefont {Sadashivaiah}}, \bibinfo {author}
  {\bibfnamefont {I.}~\bibnamefont {Sergeev}}, \bibinfo {author} {\bibfnamefont
  {O.}~\bibnamefont {Leupold}}, \bibinfo {author} {\bibfnamefont {A.~I.}\
  \bibnamefont {Chumakov}}, \bibinfo {author} {\bibfnamefont {O.}~\bibnamefont
  {Kocharovskaya}}, \emph {et~al.},\ }\href
  {https://doi.org/10.1126/sciadv.adn9825} {\bibfield  {journal} {\bibinfo
  {journal} {Sci. Adv.}\ }\textbf {\bibinfo {volume} {10}},\ \bibinfo {pages}
  {eadn9825} (\bibinfo {year} {2024})}\BibitemShut {NoStop}%
\bibitem [{\citenamefont {Haber}\ \emph {et~al.}(2019)\citenamefont {Haber},
  \citenamefont {Gollwitzer}, \citenamefont {Francoual}, \citenamefont
  {Tolkiehn}, \citenamefont {Strempfer},\ and\ \citenamefont
  {R\"ohlsberger}}]{haber2019}%
  \BibitemOpen
  \bibfield  {author} {\bibinfo {author} {\bibfnamefont {J.}~\bibnamefont
  {Haber}}, \bibinfo {author} {\bibfnamefont {J.}~\bibnamefont {Gollwitzer}},
  \bibinfo {author} {\bibfnamefont {S.}~\bibnamefont {Francoual}}, \bibinfo
  {author} {\bibfnamefont {M.}~\bibnamefont {Tolkiehn}}, \bibinfo {author}
  {\bibfnamefont {J.}~\bibnamefont {Strempfer}},\ and\ \bibinfo {author}
  {\bibfnamefont {R.}~\bibnamefont {R\"ohlsberger}},\ }\href
  {https://doi.org/10.1103/PhysRevLett.122.123608} {\bibfield  {journal}
  {\bibinfo  {journal} {Phys. Rev. Lett.}\ }\textbf {\bibinfo {volume} {122}},\
  \bibinfo {pages} {123608} (\bibinfo {year} {2019})}\BibitemShut {NoStop}%
\bibitem [{\citenamefont {Huang}\ \emph {et~al.}(2021)\citenamefont {Huang},
  \citenamefont {Kong}, \citenamefont {Li}, \citenamefont {Ma}, \citenamefont
  {Wang}, \citenamefont {Liu}, \citenamefont {Wang}, \citenamefont {Li},\ and\
  \citenamefont {Zhu}}]{huang2021}%
  \BibitemOpen
  \bibfield  {author} {\bibinfo {author} {\bibfnamefont {X.-C.}\ \bibnamefont
  {Huang}}, \bibinfo {author} {\bibfnamefont {X.-J.}\ \bibnamefont {Kong}},
  \bibinfo {author} {\bibfnamefont {T.-J.}\ \bibnamefont {Li}}, \bibinfo
  {author} {\bibfnamefont {Z.-R.}\ \bibnamefont {Ma}}, \bibinfo {author}
  {\bibfnamefont {H.-C.}\ \bibnamefont {Wang}}, \bibinfo {author}
  {\bibfnamefont {G.-C.}\ \bibnamefont {Liu}}, \bibinfo {author} {\bibfnamefont
  {Z.-S.}\ \bibnamefont {Wang}}, \bibinfo {author} {\bibfnamefont {W.-B.}\
  \bibnamefont {Li}},\ and\ \bibinfo {author} {\bibfnamefont {L.-F.}\
  \bibnamefont {Zhu}},\ }\href
  {https://doi.org/10.1103/PhysRevResearch.3.033063} {\bibfield  {journal}
  {\bibinfo  {journal} {Phys. Rev. Res.}\ }\textbf {\bibinfo {volume} {3}},\
  \bibinfo {pages} {033063} (\bibinfo {year} {2021})}\BibitemShut {NoStop}%
\bibitem [{\citenamefont {Vassholz}\ and\ \citenamefont
  {Salditt}(2021)}]{vassholz2021}%
  \BibitemOpen
  \bibfield  {author} {\bibinfo {author} {\bibfnamefont {M.}~\bibnamefont
  {Vassholz}}\ and\ \bibinfo {author} {\bibfnamefont {T.}~\bibnamefont
  {Salditt}},\ }\href {https://doi.org/10.1126/sciadv.abd5677} {\bibfield
  {journal} {\bibinfo  {journal} {Sci. Adv.}\ }\textbf {\bibinfo {volume}
  {7}},\ \bibinfo {pages} {eabd5677} (\bibinfo {year} {2021})}\BibitemShut
  {NoStop}%
\bibitem [{\citenamefont {Gu}\ \emph {et~al.}(2021)\citenamefont {Gu},
  \citenamefont {Nenov}, \citenamefont {Segatta}, \citenamefont {Garavelli},\
  and\ \citenamefont {Mukamel}}]{gu2021}%
  \BibitemOpen
  \bibfield  {author} {\bibinfo {author} {\bibfnamefont {B.}~\bibnamefont
  {Gu}}, \bibinfo {author} {\bibfnamefont {A.}~\bibnamefont {Nenov}}, \bibinfo
  {author} {\bibfnamefont {F.}~\bibnamefont {Segatta}}, \bibinfo {author}
  {\bibfnamefont {M.}~\bibnamefont {Garavelli}},\ and\ \bibinfo {author}
  {\bibfnamefont {S.}~\bibnamefont {Mukamel}},\ }\href
  {https://doi.org/10.1103/PhysRevLett.126.053201} {\bibfield  {journal}
  {\bibinfo  {journal} {Phys. Rev. Lett.}\ }\textbf {\bibinfo {volume} {126}},\
  \bibinfo {pages} {053201} (\bibinfo {year} {2021})}\BibitemShut {NoStop}%
\bibitem [{\citenamefont {Ma}\ \emph {et~al.}(2022)\citenamefont {Ma},
  \citenamefont {Huang}, \citenamefont {Li}, \citenamefont {Wang},
  \citenamefont {Liu}, \citenamefont {Wang}, \citenamefont {Li}, \citenamefont
  {Li},\ and\ \citenamefont {Zhu}}]{ma2022}%
  \BibitemOpen
  \bibfield  {author} {\bibinfo {author} {\bibfnamefont {Z.-R.}\ \bibnamefont
  {Ma}}, \bibinfo {author} {\bibfnamefont {X.-C.}\ \bibnamefont {Huang}},
  \bibinfo {author} {\bibfnamefont {T.-J.}\ \bibnamefont {Li}}, \bibinfo
  {author} {\bibfnamefont {H.-C.}\ \bibnamefont {Wang}}, \bibinfo {author}
  {\bibfnamefont {G.-C.}\ \bibnamefont {Liu}}, \bibinfo {author} {\bibfnamefont
  {Z.-S.}\ \bibnamefont {Wang}}, \bibinfo {author} {\bibfnamefont
  {B.}~\bibnamefont {Li}}, \bibinfo {author} {\bibfnamefont {W.-B.}\
  \bibnamefont {Li}},\ and\ \bibinfo {author} {\bibfnamefont {L.-F.}\
  \bibnamefont {Zhu}},\ }\href {https://doi.org/10.1103/PhysRevLett.129.213602}
  {\bibfield  {journal} {\bibinfo  {journal} {Phys. Rev. Lett.}\ }\textbf
  {\bibinfo {volume} {129}},\ \bibinfo {pages} {213602} (\bibinfo {year}
  {2022})}\BibitemShut {NoStop}%
\bibitem [{\citenamefont {Huang}\ \emph {et~al.}(2024)\citenamefont {Huang},
  \citenamefont {Li}, \citenamefont {Lima},\ and\ \citenamefont
  {Zhu}}]{huang2024}%
  \BibitemOpen
  \bibfield  {author} {\bibinfo {author} {\bibfnamefont {X.-C.}\ \bibnamefont
  {Huang}}, \bibinfo {author} {\bibfnamefont {T.-J.}\ \bibnamefont {Li}},
  \bibinfo {author} {\bibfnamefont {F.~A.}\ \bibnamefont {Lima}},\ and\
  \bibinfo {author} {\bibfnamefont {L.-F.}\ \bibnamefont {Zhu}},\ }\href
  {https://doi.org/10.1103/PhysRevA.109.033703} {\bibfield  {journal} {\bibinfo
   {journal} {Phys. Rev. A}\ }\textbf {\bibinfo {volume} {109}},\ \bibinfo
  {pages} {033703} (\bibinfo {year} {2024})}\BibitemShut {NoStop}%
\bibitem [{\citenamefont {Blatt}\ and\ \citenamefont {Roos}(2012)}]{blatt2012}%
  \BibitemOpen
  \bibfield  {author} {\bibinfo {author} {\bibfnamefont {R.}~\bibnamefont
  {Blatt}}\ and\ \bibinfo {author} {\bibfnamefont {C.~F.}\ \bibnamefont
  {Roos}},\ }\href@noop {} {\bibfield  {journal} {\bibinfo  {journal} {Nature
  Physics}\ }\textbf {\bibinfo {volume} {8}},\ \bibinfo {pages} {277} (\bibinfo
  {year} {2012})}\BibitemShut {NoStop}%
\bibitem [{\citenamefont {Blais}\ \emph {et~al.}(2021)\citenamefont {Blais},
  \citenamefont {Grimsmo}, \citenamefont {Girvin},\ and\ \citenamefont
  {Wallraff}}]{blais2021}%
  \BibitemOpen
  \bibfield  {author} {\bibinfo {author} {\bibfnamefont {A.}~\bibnamefont
  {Blais}}, \bibinfo {author} {\bibfnamefont {A.~L.}\ \bibnamefont {Grimsmo}},
  \bibinfo {author} {\bibfnamefont {S.~M.}\ \bibnamefont {Girvin}},\ and\
  \bibinfo {author} {\bibfnamefont {A.}~\bibnamefont {Wallraff}},\ }\href
  {https://doi.org/10.1103/RevModPhys.93.025005} {\bibfield  {journal}
  {\bibinfo  {journal} {Rev. Mod. Phys.}\ }\textbf {\bibinfo {volume} {93}},\
  \bibinfo {pages} {025005} (\bibinfo {year} {2021})}\BibitemShut {NoStop}%
\bibitem [{\citenamefont {Gel'mukhanov}\ \emph {et~al.}(2021)\citenamefont
  {Gel'mukhanov}, \citenamefont {Odelius}, \citenamefont {Polyutov},
  \citenamefont {F\"ohlisch},\ and\ \citenamefont {Kimberg}}]{gelmukhanov2021}%
  \BibitemOpen
  \bibfield  {author} {\bibinfo {author} {\bibfnamefont {F.}~\bibnamefont
  {Gel'mukhanov}}, \bibinfo {author} {\bibfnamefont {M.}~\bibnamefont
  {Odelius}}, \bibinfo {author} {\bibfnamefont {S.~P.}\ \bibnamefont
  {Polyutov}}, \bibinfo {author} {\bibfnamefont {A.}~\bibnamefont
  {F\"ohlisch}},\ and\ \bibinfo {author} {\bibfnamefont {V.}~\bibnamefont
  {Kimberg}},\ }\href {https://doi.org/10.1103/RevModPhys.93.035001} {\bibfield
   {journal} {\bibinfo  {journal} {Rev. Mod. Phys.}\ }\textbf {\bibinfo
  {volume} {93}},\ \bibinfo {pages} {035001} (\bibinfo {year}
  {2021})}\BibitemShut {NoStop}%
\bibitem [{\citenamefont {De~Groot}\ and\ \citenamefont
  {Kotani}(2008)}]{groot2008}%
  \BibitemOpen
  \bibfield  {author} {\bibinfo {author} {\bibfnamefont {F.}~\bibnamefont
  {De~Groot}}\ and\ \bibinfo {author} {\bibfnamefont {A.}~\bibnamefont
  {Kotani}},\ }\href@noop {} {\emph {\bibinfo {title} {Core level spectroscopy
  of solids}}}\ (\bibinfo  {publisher} {CRC press, Talylor \& Francis},\
  \bibinfo {year} {2008})\BibitemShut {NoStop}%
\bibitem [{\citenamefont {Kotani}\ and\ \citenamefont
  {Shin}(2001)}]{kotani2001}%
  \BibitemOpen
  \bibfield  {author} {\bibinfo {author} {\bibfnamefont {A.}~\bibnamefont
  {Kotani}}\ and\ \bibinfo {author} {\bibfnamefont {S.}~\bibnamefont {Shin}},\
  }\href {https://doi.org/10.1103/RevModPhys.73.203} {\bibfield  {journal}
  {\bibinfo  {journal} {Rev. Mod. Phys.}\ }\textbf {\bibinfo {volume} {73}},\
  \bibinfo {pages} {203} (\bibinfo {year} {2001})}\BibitemShut {NoStop}%
\bibitem [{\citenamefont {Ament}\ \emph {et~al.}(2011)\citenamefont {Ament},
  \citenamefont {van Veenendaal}, \citenamefont {Devereaux}, \citenamefont
  {Hill},\ and\ \citenamefont {van~den Brink}}]{ament2011}%
  \BibitemOpen
  \bibfield  {author} {\bibinfo {author} {\bibfnamefont {L.~J.~P.}\
  \bibnamefont {Ament}}, \bibinfo {author} {\bibfnamefont {M.}~\bibnamefont
  {van Veenendaal}}, \bibinfo {author} {\bibfnamefont {T.~P.}\ \bibnamefont
  {Devereaux}}, \bibinfo {author} {\bibfnamefont {J.~P.}\ \bibnamefont
  {Hill}},\ and\ \bibinfo {author} {\bibfnamefont {J.}~\bibnamefont {van~den
  Brink}},\ }\href {https://doi.org/10.1103/RevModPhys.83.705} {\bibfield
  {journal} {\bibinfo  {journal} {Rev. Mod. Phys.}\ }\textbf {\bibinfo {volume}
  {83}},\ \bibinfo {pages} {705} (\bibinfo {year} {2011})}\BibitemShut
  {NoStop}%
\bibitem [{\citenamefont {Gel'mukhanov}\ and\ \citenamefont
  {{\AA}gren}(1999)}]{gelmukhanov1999}%
  \BibitemOpen
  \bibfield  {author} {\bibinfo {author} {\bibfnamefont {F.}~\bibnamefont
  {Gel'mukhanov}}\ and\ \bibinfo {author} {\bibfnamefont {H.}~\bibnamefont
  {{\AA}gren}},\ }\href {https://doi.org/10.1016/S0370-1573(99)00003-4}
  {\bibfield  {journal} {\bibinfo  {journal} {Phys. Rep.}\ }\textbf {\bibinfo
  {volume} {312}},\ \bibinfo {pages} {87} (\bibinfo {year} {1999})}\BibitemShut
  {NoStop}%
\bibitem [{\citenamefont {Ghiringhelli}\ \emph {et~al.}(2005)\citenamefont
  {Ghiringhelli}, \citenamefont {Matsubara}, \citenamefont {Dallera},
  \citenamefont {Fracassi}, \citenamefont {Gusmeroli}, \citenamefont
  {Piazzalunga}, \citenamefont {Tagliaferri}, \citenamefont {Brookes},
  \citenamefont {Kotani},\ and\ \citenamefont {Braicovich}}]{ghiringhelli2005}%
  \BibitemOpen
  \bibfield  {author} {\bibinfo {author} {\bibfnamefont {G.}~\bibnamefont
  {Ghiringhelli}}, \bibinfo {author} {\bibfnamefont {M.}~\bibnamefont
  {Matsubara}}, \bibinfo {author} {\bibfnamefont {C.}~\bibnamefont {Dallera}},
  \bibinfo {author} {\bibfnamefont {F.}~\bibnamefont {Fracassi}}, \bibinfo
  {author} {\bibfnamefont {R.}~\bibnamefont {Gusmeroli}}, \bibinfo {author}
  {\bibfnamefont {A.}~\bibnamefont {Piazzalunga}}, \bibinfo {author}
  {\bibfnamefont {A.}~\bibnamefont {Tagliaferri}}, \bibinfo {author}
  {\bibfnamefont {N.~B.}\ \bibnamefont {Brookes}}, \bibinfo {author}
  {\bibfnamefont {A.}~\bibnamefont {Kotani}},\ and\ \bibinfo {author}
  {\bibfnamefont {L.}~\bibnamefont {Braicovich}},\ }\href
  {https://doi.org/10.1088/0953-8984/17/35/007} {\bibfield  {journal} {\bibinfo
   {journal} {J. Phys. Condens. Matter}\ }\textbf {\bibinfo {volume} {17}},\
  \bibinfo {pages} {5397} (\bibinfo {year} {2005})}\BibitemShut {NoStop}%
\bibitem [{\citenamefont {Baker}\ \emph {et~al.}(2017)\citenamefont {Baker},
  \citenamefont {Mara}, \citenamefont {Yan}, \citenamefont {Hodgson},
  \citenamefont {Hedman},\ and\ \citenamefont {Solomon}}]{baker2017}%
  \BibitemOpen
  \bibfield  {author} {\bibinfo {author} {\bibfnamefont {M.~L.}\ \bibnamefont
  {Baker}}, \bibinfo {author} {\bibfnamefont {M.~W.}\ \bibnamefont {Mara}},
  \bibinfo {author} {\bibfnamefont {J.~J.}\ \bibnamefont {Yan}}, \bibinfo
  {author} {\bibfnamefont {K.~O.}\ \bibnamefont {Hodgson}}, \bibinfo {author}
  {\bibfnamefont {B.}~\bibnamefont {Hedman}},\ and\ \bibinfo {author}
  {\bibfnamefont {E.~I.}\ \bibnamefont {Solomon}},\ }\href
  {https://doi.org/https://doi.org/10.1016/j.ccr.2017.02.004} {\bibfield
  {journal} {\bibinfo  {journal} {Coord. Chem. Rev.}\ }\textbf {\bibinfo
  {volume} {345}},\ \bibinfo {pages} {182} (\bibinfo {year}
  {2017})}\BibitemShut {NoStop}%
\bibitem [{\citenamefont {Heeg}\ \emph {et~al.}(2013)\citenamefont {Heeg},
  \citenamefont {Wille}, \citenamefont {Schlage}, \citenamefont {Guryeva},
  \citenamefont {Schumacher}, \citenamefont {Uschmann}, \citenamefont
  {Schulze}, \citenamefont {Marx}, \citenamefont {K\"ampfer}, \citenamefont
  {Paulus}, \citenamefont {R\"ohlsberger},\ and\ \citenamefont
  {Evers}}]{heeg2013prl}%
  \BibitemOpen
  \bibfield  {author} {\bibinfo {author} {\bibfnamefont {K.~P.}\ \bibnamefont
  {Heeg}}, \bibinfo {author} {\bibfnamefont {H.-C.}\ \bibnamefont {Wille}},
  \bibinfo {author} {\bibfnamefont {K.}~\bibnamefont {Schlage}}, \bibinfo
  {author} {\bibfnamefont {T.}~\bibnamefont {Guryeva}}, \bibinfo {author}
  {\bibfnamefont {D.}~\bibnamefont {Schumacher}}, \bibinfo {author}
  {\bibfnamefont {I.}~\bibnamefont {Uschmann}}, \bibinfo {author}
  {\bibfnamefont {K.~S.}\ \bibnamefont {Schulze}}, \bibinfo {author}
  {\bibfnamefont {B.}~\bibnamefont {Marx}}, \bibinfo {author} {\bibfnamefont
  {T.}~\bibnamefont {K\"ampfer}}, \bibinfo {author} {\bibfnamefont {G.~G.}\
  \bibnamefont {Paulus}}, \bibinfo {author} {\bibfnamefont {R.}~\bibnamefont
  {R\"ohlsberger}},\ and\ \bibinfo {author} {\bibfnamefont {J.}~\bibnamefont
  {Evers}},\ }\href {https://doi.org/10.1103/PhysRevLett.111.073601} {\bibfield
   {journal} {\bibinfo  {journal} {Phys. Rev. Lett.}\ }\textbf {\bibinfo
  {volume} {111}},\ \bibinfo {pages} {073601} (\bibinfo {year}
  {2013})}\BibitemShut {NoStop}%
\bibitem [{\citenamefont {Kong}\ and\ \citenamefont
  {P\'alffy}(2016)}]{kong2016}%
  \BibitemOpen
  \bibfield  {author} {\bibinfo {author} {\bibfnamefont {X.}~\bibnamefont
  {Kong}}\ and\ \bibinfo {author} {\bibfnamefont {A.}~\bibnamefont
  {P\'alffy}},\ }\href {https://doi.org/10.1103/PhysRevLett.116.197402}
  {\bibfield  {journal} {\bibinfo  {journal} {Phys. Rev. Lett.}\ }\textbf
  {\bibinfo {volume} {116}},\ \bibinfo {pages} {197402} (\bibinfo {year}
  {2016})}\BibitemShut {NoStop}%
\bibitem [{\citenamefont {Sch{\"u}lke}(2007)}]{schulke2007}%
  \BibitemOpen
  \bibfield  {author} {\bibinfo {author} {\bibfnamefont {W.}~\bibnamefont
  {Sch{\"u}lke}},\ }\href@noop {} {\emph {\bibinfo {title} {Electron dynamics
  by inelastic X-ray scattering}}},\ Vol.~\bibinfo {volume} {7}\ (\bibinfo
  {publisher} {Oxford University Press},\ \bibinfo {year} {2007})\BibitemShut
  {NoStop}%
\bibitem [{\citenamefont {Heeg}\ and\ \citenamefont
  {Evers}(2013)}]{heeg2013pra}%
  \BibitemOpen
  \bibfield  {author} {\bibinfo {author} {\bibfnamefont {K.~P.}\ \bibnamefont
  {Heeg}}\ and\ \bibinfo {author} {\bibfnamefont {J.}~\bibnamefont {Evers}},\
  }\href {https://doi.org/10.1103/PhysRevA.88.043828} {\bibfield  {journal}
  {\bibinfo  {journal} {Phys. Rev. A}\ }\textbf {\bibinfo {volume} {88}},\
  \bibinfo {pages} {043828} (\bibinfo {year} {2013})}\BibitemShut {NoStop}%
\bibitem [{\citenamefont {Heeg}\ and\ \citenamefont
  {Evers}(2015)}]{heeg2015pra}%
  \BibitemOpen
  \bibfield  {author} {\bibinfo {author} {\bibfnamefont {K.~P.}\ \bibnamefont
  {Heeg}}\ and\ \bibinfo {author} {\bibfnamefont {J.}~\bibnamefont {Evers}},\
  }\href {https://doi.org/10.1103/PhysRevA.91.063803} {\bibfield  {journal}
  {\bibinfo  {journal} {Phys. Rev. A}\ }\textbf {\bibinfo {volume} {91}},\
  \bibinfo {pages} {063803} (\bibinfo {year} {2015})}\BibitemShut {NoStop}%
\bibitem [{\citenamefont {Kong}\ \emph {et~al.}(2020)\citenamefont {Kong},
  \citenamefont {Chang},\ and\ \citenamefont {P\'alffy}}]{kong2020}%
  \BibitemOpen
  \bibfield  {author} {\bibinfo {author} {\bibfnamefont {X.}~\bibnamefont
  {Kong}}, \bibinfo {author} {\bibfnamefont {D.~E.}\ \bibnamefont {Chang}},\
  and\ \bibinfo {author} {\bibfnamefont {A.}~\bibnamefont {P\'alffy}},\ }\href
  {https://doi.org/10.1103/PhysRevA.102.033710} {\bibfield  {journal} {\bibinfo
   {journal} {Phys. Rev. A}\ }\textbf {\bibinfo {volume} {102}},\ \bibinfo
  {pages} {033710} (\bibinfo {year} {2020})}\BibitemShut {NoStop}%
\bibitem [{\citenamefont {Lentrodt}\ \emph {et~al.}(2020)\citenamefont
  {Lentrodt}, \citenamefont {Heeg}, \citenamefont {Keitel},\ and\ \citenamefont
  {Evers}}]{lentrodt2020prr}%
  \BibitemOpen
  \bibfield  {author} {\bibinfo {author} {\bibfnamefont {D.}~\bibnamefont
  {Lentrodt}}, \bibinfo {author} {\bibfnamefont {K.~P.}\ \bibnamefont {Heeg}},
  \bibinfo {author} {\bibfnamefont {C.~H.}\ \bibnamefont {Keitel}},\ and\
  \bibinfo {author} {\bibfnamefont {J.}~\bibnamefont {Evers}},\ }\href
  {https://doi.org/10.1103/PhysRevResearch.2.023396} {\bibfield  {journal}
  {\bibinfo  {journal} {Phys. Rev. Res.}\ }\textbf {\bibinfo {volume} {2}},\
  \bibinfo {pages} {023396} (\bibinfo {year} {2020})}\BibitemShut {NoStop}%
\bibitem [{\citenamefont {De~Groot}(2001)}]{groot2001high}%
  \BibitemOpen
  \bibfield  {author} {\bibinfo {author} {\bibfnamefont {F.}~\bibnamefont
  {De~Groot}},\ }\href {https://doi.org/10.1021/cr9900681} {\bibfield
  {journal} {\bibinfo  {journal} {Chem. Rev.}\ }\textbf {\bibinfo {volume}
  {101}},\ \bibinfo {pages} {1779} (\bibinfo {year} {2001})}\BibitemShut
  {NoStop}%
\bibitem [{\citenamefont {B\l{}achucki}\ \emph {et~al.}(2014)\citenamefont
  {B\l{}achucki}, \citenamefont {Szlachetko}, \citenamefont {Hoszowska},
  \citenamefont {Dousse}, \citenamefont {Kayser}, \citenamefont {Nachtegaal},\
  and\ \citenamefont {S\'a}}]{blachucki2014}%
  \BibitemOpen
  \bibfield  {author} {\bibinfo {author} {\bibfnamefont {W.}~\bibnamefont
  {B\l{}achucki}}, \bibinfo {author} {\bibfnamefont {J.}~\bibnamefont
  {Szlachetko}}, \bibinfo {author} {\bibfnamefont {J.}~\bibnamefont
  {Hoszowska}}, \bibinfo {author} {\bibfnamefont {J.-C.}\ \bibnamefont
  {Dousse}}, \bibinfo {author} {\bibfnamefont {Y.}~\bibnamefont {Kayser}},
  \bibinfo {author} {\bibfnamefont {M.}~\bibnamefont {Nachtegaal}},\ and\
  \bibinfo {author} {\bibfnamefont {J.}~\bibnamefont {S\'a}},\ }\href
  {https://doi.org/10.1103/PhysRevLett.112.173003} {\bibfield  {journal}
  {\bibinfo  {journal} {Phys. Rev. Lett.}\ }\textbf {\bibinfo {volume} {112}},\
  \bibinfo {pages} {173003} (\bibinfo {year} {2014})}\BibitemShut {NoStop}%
\bibitem [{\citenamefont {Bauer}(2014)}]{bauer2014}%
  \BibitemOpen
  \bibfield  {author} {\bibinfo {author} {\bibfnamefont {M.}~\bibnamefont
  {Bauer}},\ }\href {https://doi.org/10.1039/C4CP00904E} {\bibfield  {journal}
  {\bibinfo  {journal} {Phys. Chem. Chem. Phys.}\ }\textbf {\bibinfo {volume}
  {16}},\ \bibinfo {pages} {13827} (\bibinfo {year} {2014})}\BibitemShut
  {NoStop}%
\bibitem [{\citenamefont {de~Groot}\ \emph {et~al.}(2005)\citenamefont
  {de~Groot}, \citenamefont {Glatzel}, \citenamefont {Bergmann}, \citenamefont
  {van Aken}, \citenamefont {Barrea}, \citenamefont {Klemme}, \citenamefont
  {Hävecker}, \citenamefont {Knop-Gericke}, \citenamefont {Heijboer},\ and\
  \citenamefont {Weckhuysen}}]{groot2005}%
  \BibitemOpen
  \bibfield  {author} {\bibinfo {author} {\bibfnamefont {F.~M.~F.}\
  \bibnamefont {de~Groot}}, \bibinfo {author} {\bibfnamefont {P.}~\bibnamefont
  {Glatzel}}, \bibinfo {author} {\bibfnamefont {U.}~\bibnamefont {Bergmann}},
  \bibinfo {author} {\bibfnamefont {P.~A.}\ \bibnamefont {van Aken}}, \bibinfo
  {author} {\bibfnamefont {R.~A.}\ \bibnamefont {Barrea}}, \bibinfo {author}
  {\bibfnamefont {S.}~\bibnamefont {Klemme}}, \bibinfo {author} {\bibfnamefont
  {M.}~\bibnamefont {Hävecker}}, \bibinfo {author} {\bibfnamefont
  {A.}~\bibnamefont {Knop-Gericke}}, \bibinfo {author} {\bibfnamefont {W.~M.}\
  \bibnamefont {Heijboer}},\ and\ \bibinfo {author} {\bibfnamefont {B.~M.}\
  \bibnamefont {Weckhuysen}},\ }\href {https://doi.org/10.1021/jp054006s}
  {\bibfield  {journal} {\bibinfo  {journal} {J. Phys. Chem. B}\ }\textbf
  {\bibinfo {volume} {109}},\ \bibinfo {pages} {20751} (\bibinfo {year}
  {2005})}\BibitemShut {NoStop}%
\bibitem [{sup()}]{supp2024}%
  \BibitemOpen
  \href@noop {} {}\bibinfo {note} {See Supplemental Material for
  details.}\BibitemShut {Stop}%
\bibitem [{\citenamefont {Zhao}\ \emph {et~al.}(2025)\citenamefont {Zhao},
  \citenamefont {Wang}, \citenamefont {Wang}, \citenamefont {Su}, \citenamefont
  {Ma}, \citenamefont {Huang},\ and\ \citenamefont {Zhu}}]{zhao2025}%
  \BibitemOpen
  \bibfield  {author} {\bibinfo {author} {\bibfnamefont {Z.~Q.}\ \bibnamefont
  {Zhao}}, \bibinfo {author} {\bibfnamefont {S.~X.}\ \bibnamefont {Wang}},
  \bibinfo {author} {\bibfnamefont {X.~Y.}\ \bibnamefont {Wang}}, \bibinfo
  {author} {\bibfnamefont {Y.}~\bibnamefont {Su}}, \bibinfo {author}
  {\bibfnamefont {Z.~R.}\ \bibnamefont {Ma}}, \bibinfo {author} {\bibfnamefont
  {X.~C.}\ \bibnamefont {Huang}},\ and\ \bibinfo {author} {\bibfnamefont
  {L.~F.}\ \bibnamefont {Zhu}},\ }\href
  {https://doi.org/10.7498/aps.74.20250659} {\bibfield  {journal} {\bibinfo
  {journal} {Acta Phys. Sin.}\ }\textbf {\bibinfo {volume} {74}},\ \bibinfo
  {pages} {183201} (\bibinfo {year} {2025})}\BibitemShut {NoStop}%
\bibitem [{\citenamefont {Johann}(1931)}]{johann1931}%
  \BibitemOpen
  \bibfield  {author} {\bibinfo {author} {\bibfnamefont {H.~H.}\ \bibnamefont
  {Johann}},\ }\href@noop {} {\bibfield  {journal} {\bibinfo  {journal}
  {Zeitschrift f{\"u}r Physik}\ }\textbf {\bibinfo {volume} {69}},\ \bibinfo
  {pages} {185} (\bibinfo {year} {1931})}\BibitemShut {NoStop}%
\bibitem [{\citenamefont {DuMond}(1947)}]{dumond1947}%
  \BibitemOpen
  \bibfield  {author} {\bibinfo {author} {\bibfnamefont {J.~W.~M.}\
  \bibnamefont {DuMond}},\ }\href {https://doi.org/10.1063/1.1741017}
  {\bibfield  {journal} {\bibinfo  {journal} {Rev. Sci. Instrum.}\ }\textbf
  {\bibinfo {volume} {18}},\ \bibinfo {pages} {626} (\bibinfo {year}
  {1947})}\BibitemShut {NoStop}%
\bibitem [{\citenamefont {V.~H{\'a}mos}(1933)}]{hamos1933}%
  \BibitemOpen
  \bibfield  {author} {\bibinfo {author} {\bibfnamefont {L.}~\bibnamefont
  {V.~H{\'a}mos}},\ }\href@noop {} {\bibfield  {journal} {\bibinfo  {journal}
  {Ann. Phys. (Leipzig)}\ }\textbf {\bibinfo {volume} {409}},\ \bibinfo {pages}
  {716} (\bibinfo {year} {1933})}\BibitemShut {NoStop}%
\bibitem [{\citenamefont {Wach}\ \emph {et~al.}(2020)\citenamefont {Wach},
  \citenamefont {S{\'{a}}},\ and\ \citenamefont {Szlachetko}}]{wach2020}%
  \BibitemOpen
  \bibfield  {author} {\bibinfo {author} {\bibfnamefont {A.}~\bibnamefont
  {Wach}}, \bibinfo {author} {\bibfnamefont {J.}~\bibnamefont {S{\'{a}}}},\
  and\ \bibinfo {author} {\bibfnamefont {J.}~\bibnamefont {Szlachetko}},\
  }\href {https://doi.org/10.1107/S1600577520003690} {\bibfield  {journal}
  {\bibinfo  {journal} {J. Synchrotron Rad.}\ }\textbf {\bibinfo {volume}
  {27}},\ \bibinfo {pages} {689} (\bibinfo {year} {2020})}\BibitemShut
  {NoStop}%
\bibitem [{\citenamefont {Ablett}\ \emph {et~al.}(2019)\citenamefont {Ablett},
  \citenamefont {Prieur}, \citenamefont {Céolin}, \citenamefont
  {Lassalle-Kaiser}, \citenamefont {Lebert}, \citenamefont {Sauvage},
  \citenamefont {Moreno}, \citenamefont {Bac}, \citenamefont {Balédent},
  \citenamefont {Ovono}, \citenamefont {Morand}, \citenamefont {Gélebart},
  \citenamefont {Shukla},\ and\ \citenamefont {Rueff}}]{ablett2019galaxies}%
  \BibitemOpen
  \bibfield  {author} {\bibinfo {author} {\bibfnamefont {J.~M.}\ \bibnamefont
  {Ablett}}, \bibinfo {author} {\bibfnamefont {D.}~\bibnamefont {Prieur}},
  \bibinfo {author} {\bibfnamefont {D.}~\bibnamefont {Céolin}}, \bibinfo
  {author} {\bibfnamefont {B.}~\bibnamefont {Lassalle-Kaiser}}, \bibinfo
  {author} {\bibfnamefont {B.}~\bibnamefont {Lebert}}, \bibinfo {author}
  {\bibfnamefont {M.}~\bibnamefont {Sauvage}}, \bibinfo {author} {\bibfnamefont
  {T.}~\bibnamefont {Moreno}}, \bibinfo {author} {\bibfnamefont
  {S.}~\bibnamefont {Bac}}, \bibinfo {author} {\bibfnamefont {V.}~\bibnamefont
  {Balédent}}, \bibinfo {author} {\bibfnamefont {A.}~\bibnamefont {Ovono}},
  \bibinfo {author} {\bibfnamefont {M.}~\bibnamefont {Morand}}, \bibinfo
  {author} {\bibfnamefont {F.}~\bibnamefont {Gélebart}}, \bibinfo {author}
  {\bibfnamefont {A.}~\bibnamefont {Shukla}},\ and\ \bibinfo {author}
  {\bibfnamefont {J.-P.}\ \bibnamefont {Rueff}},\ }\href
  {https://doi.org/https://doi.org/10.1107/S160057751801559X} {\bibfield
  {journal} {\bibinfo  {journal} {J. Synchrotron Rad.}\ }\textbf {\bibinfo
  {volume} {26}},\ \bibinfo {pages} {263} (\bibinfo {year} {2019})}\BibitemShut
  {NoStop}%
\bibitem [{\citenamefont {Zegenhagen}\ and\ \citenamefont
  {Kazimirov}(2013)}]{zegenhagen2013}%
  \BibitemOpen
  \bibfield  {author} {\bibinfo {author} {\bibfnamefont {J.}~\bibnamefont
  {Zegenhagen}}\ and\ \bibinfo {author} {\bibfnamefont {A.}~\bibnamefont
  {Kazimirov}},\ }\href@noop {} {\emph {\bibinfo {title} {X-ray Standing Wave
  Technique, The: Principles And Applications}}},\ Vol.~\bibinfo {volume} {7}\
  (\bibinfo  {publisher} {World Scientific},\ \bibinfo {year}
  {2013})\BibitemShut {NoStop}%
\bibitem [{\citenamefont {Andreji\ifmmode~\acute{c}\else \'{c}\fi{}}\ and\
  \citenamefont {P\'alffy}(2021)}]{andreji2021}%
  \BibitemOpen
  \bibfield  {author} {\bibinfo {author} {\bibfnamefont {P.}~\bibnamefont
  {Andreji\ifmmode~\acute{c}\else \'{c}\fi{}}}\ and\ \bibinfo {author}
  {\bibfnamefont {A.}~\bibnamefont {P\'alffy}},\ }\href
  {https://doi.org/10.1103/PhysRevA.104.033702} {\bibfield  {journal} {\bibinfo
   {journal} {Phys. Rev. A}\ }\textbf {\bibinfo {volume} {104}},\ \bibinfo
  {pages} {033702} (\bibinfo {year} {2021})}\BibitemShut {NoStop}%
\bibitem [{\citenamefont {Haber}\ \emph {et~al.}(2016)\citenamefont {Haber},
  \citenamefont {Schulze}, \citenamefont {Schlage}, \citenamefont {Loetzsch},
  \citenamefont {Bocklage}, \citenamefont {Gurieva}, \citenamefont {Bernhardt},
  \citenamefont {Wille}, \citenamefont {R{\"u}ffer}, \citenamefont {Uschmann},
  \citenamefont {Paulus},\ and\ \citenamefont {Röhlsberger}}]{haber2016}%
  \BibitemOpen
  \bibfield  {author} {\bibinfo {author} {\bibfnamefont {J.}~\bibnamefont
  {Haber}}, \bibinfo {author} {\bibfnamefont {K.~S.}\ \bibnamefont {Schulze}},
  \bibinfo {author} {\bibfnamefont {K.}~\bibnamefont {Schlage}}, \bibinfo
  {author} {\bibfnamefont {R.}~\bibnamefont {Loetzsch}}, \bibinfo {author}
  {\bibfnamefont {L.}~\bibnamefont {Bocklage}}, \bibinfo {author}
  {\bibfnamefont {T.}~\bibnamefont {Gurieva}}, \bibinfo {author} {\bibfnamefont
  {H.}~\bibnamefont {Bernhardt}}, \bibinfo {author} {\bibfnamefont {H.-C.}\
  \bibnamefont {Wille}}, \bibinfo {author} {\bibfnamefont {R.}~\bibnamefont
  {R{\"u}ffer}}, \bibinfo {author} {\bibfnamefont {I.}~\bibnamefont
  {Uschmann}}, \bibinfo {author} {\bibfnamefont {G.~G.}\ \bibnamefont
  {Paulus}},\ and\ \bibinfo {author} {\bibfnamefont {R.}~\bibnamefont
  {Röhlsberger}},\ }\href {https://doi.org/10.1038/nphoton.2016.77} {\bibfield
   {journal} {\bibinfo  {journal} {Nat. Photon.}\ }\textbf {\bibinfo {volume}
  {10}},\ \bibinfo {pages} {445} (\bibinfo {year} {2016})}\BibitemShut
  {NoStop}%
\bibitem [{\citenamefont {R{\"o}hlsberger}(2004)}]{rohlsberger2004}%
  \BibitemOpen
  \bibfield  {author} {\bibinfo {author} {\bibfnamefont {R.}~\bibnamefont
  {R{\"o}hlsberger}},\ }\href@noop {} {\emph {\bibinfo {title} {Nuclear
  condensed matter physics with synchrotron radiation: Basic principles,
  methodology and applications}}}\ (\bibinfo  {publisher} {Springer Science \&
  Business Media},\ \bibinfo {year} {2004})\BibitemShut {NoStop}%
\bibitem [{\citenamefont {Liu}\ \emph {et~al.}(2023)\citenamefont {Liu},
  \citenamefont {Grech}, \citenamefont {Guetg}, \citenamefont {Karabekyan},
  \citenamefont {Kocharyan}, \citenamefont {Kujala}, \citenamefont {Lechner},
  \citenamefont {Long}, \citenamefont {Mirian}, \citenamefont {Qin} \emph
  {et~al.}}]{liu2023}%
  \BibitemOpen
  \bibfield  {author} {\bibinfo {author} {\bibfnamefont {S.}~\bibnamefont
  {Liu}}, \bibinfo {author} {\bibfnamefont {C.}~\bibnamefont {Grech}}, \bibinfo
  {author} {\bibfnamefont {M.}~\bibnamefont {Guetg}}, \bibinfo {author}
  {\bibfnamefont {S.}~\bibnamefont {Karabekyan}}, \bibinfo {author}
  {\bibfnamefont {V.}~\bibnamefont {Kocharyan}}, \bibinfo {author}
  {\bibfnamefont {N.}~\bibnamefont {Kujala}}, \bibinfo {author} {\bibfnamefont
  {C.}~\bibnamefont {Lechner}}, \bibinfo {author} {\bibfnamefont
  {T.}~\bibnamefont {Long}}, \bibinfo {author} {\bibfnamefont {N.}~\bibnamefont
  {Mirian}}, \bibinfo {author} {\bibfnamefont {W.}~\bibnamefont {Qin}}, \emph
  {et~al.},\ }\href
  {https://doi.org/https://doi.org/10.1038/s41566-023-01305-x} {\bibfield
  {journal} {\bibinfo  {journal} {Nat. Photon.}\ }\textbf {\bibinfo {volume}
  {17}},\ \bibinfo {pages} {984} (\bibinfo {year} {2023})}\BibitemShut
  {NoStop}%
\bibitem [{\citenamefont {Yan}\ \emph {et~al.}(2024)\citenamefont {Yan},
  \citenamefont {Qin}, \citenamefont {Chen}, \citenamefont {Decking},
  \citenamefont {Dijkstal}, \citenamefont {Guetg}, \citenamefont {Inoue},
  \citenamefont {Kujala}, \citenamefont {Liu}, \citenamefont {Long} \emph
  {et~al.}}]{yan2024}%
  \BibitemOpen
  \bibfield  {author} {\bibinfo {author} {\bibfnamefont {J.}~\bibnamefont
  {Yan}}, \bibinfo {author} {\bibfnamefont {W.}~\bibnamefont {Qin}}, \bibinfo
  {author} {\bibfnamefont {Y.}~\bibnamefont {Chen}}, \bibinfo {author}
  {\bibfnamefont {W.}~\bibnamefont {Decking}}, \bibinfo {author} {\bibfnamefont
  {P.}~\bibnamefont {Dijkstal}}, \bibinfo {author} {\bibfnamefont
  {M.}~\bibnamefont {Guetg}}, \bibinfo {author} {\bibfnamefont
  {I.}~\bibnamefont {Inoue}}, \bibinfo {author} {\bibfnamefont
  {N.}~\bibnamefont {Kujala}}, \bibinfo {author} {\bibfnamefont
  {S.}~\bibnamefont {Liu}}, \bibinfo {author} {\bibfnamefont {T.}~\bibnamefont
  {Long}}, \emph {et~al.},\ }\href
  {https://doi.org/https://doi.org/10.1038/s41566-024-01566-0} {\bibfield
  {journal} {\bibinfo  {journal} {Nat. Photon.}\ ,\ \bibinfo {pages} {1}}
  (\bibinfo {year} {2024})}\BibitemShut {NoStop}%
\bibitem [{\citenamefont {Rohringer}(2019)}]{rohringer2019}%
  \BibitemOpen
  \bibfield  {author} {\bibinfo {author} {\bibfnamefont {N.}~\bibnamefont
  {Rohringer}},\ }\href {https://doi.org/http://doi.org/10.1098/rsta.2017.0471}
  {\bibfield  {journal} {\bibinfo  {journal} {Phil. Trans. R. Soc. A}\ }\textbf
  {\bibinfo {volume} {377}},\ \bibinfo {pages} {20170471} (\bibinfo {year}
  {2019})}\BibitemShut {NoStop}%
\end{thebibliography}%

\end{document}


\title{Supplementary Material for: \\Cavity Controls Core-to-Core Resonant Inelastic X-ray Scattering}

\author{S.-X. Wang\orcidlink{0000-0002-6305-3762}}
\affiliation{Department of Modern Physics, University of Science and Technology of China, Hefei, Anhui 230026, China}
\affiliation{I. Physikalisches Institut, Justus-Liebig-Universität Gießen und Helmholtz Forschungsakademie Hessen für 
	FAIR (HFHF) GSI Helmholtzzentrum für Schwerionenforschung Campus Gießen, Heinrich-Buff-Ring 16, 35392 Gießen, Germany}

\author{Z.-Q. Zhao}\author{X.-Y. Wang}\author{T.-J. Li\orcidlink{0000-0002-4391-5257}}\author{Y. Su}
\affiliation{Department of Modern Physics, University of Science and Technology of China, Hefei, Anhui 230026, China}

\author{Y. Uemura\orcidlink{0000-0003-3164-7168}}\author{F. Alves Lima\orcidlink{0000-0001-8106-2892}}
\affiliation{European XFEL, 22869 Schenefeld, Germany}

\author{A. Khadiev\orcidlink{0000-0001-7577-2855}}\author{B.-H. Wang\orcidlink{0000-0002-1223-503X}}
\affiliation{Deutsches Elektronen-Synchrotron DESY, 22607 Hamburg, Germany}

\author{J. M. Ablett\orcidlink{0000-0003-2887-2903}}
\affiliation{Synchrotron SOLEIL, L’Orme des Merisiers, Départementale 128, 91190 Saint-Aubin, France}

\author{J-P. Rueff\orcidlink{0000-0003-3594-918X}}
\affiliation{Synchrotron SOLEIL, L’Orme des Merisiers, Départementale 128, 91190 Saint-Aubin, France}
\affiliation{Sorbonne Université, CNRS, Laboratoire de Chimie Physique – Matière et Rayonnement, LCPMR, F-75005 Paris, France}

\author{H.-C. Wang}\author{O. J. L. Fox\orcidlink{0000-0001-5224-7062}}
\affiliation{Diamond Light Source, Harwell Science and Innovation Campus, Didcot, Oxfordshire, OX11 0DE, United Kingdom}

\author{W.-B. Li}
\affiliation{MOE Key Laboratory of Advanced Micro-Structured Materials, Institute of Precision Optical Engineering (IPOE), School of Physics Science and 
Engineering, Tongji University, Shanghai 200092, China}

\author{L.-F. Zhu\orcidlink{0000-0002-5771-0471}}
\affiliation{Department of Modern Physics, University of Science and Technology of China, Hefei, Anhui 230026, China}

\author{X.-C. Huang\orcidlink{0000-0002-9140-6369}}\email[Corresponding author: ]{xinchao.huang@xfel.eu}
\affiliation{European XFEL, 22869 Schenefeld, Germany}

\date{\today}

\maketitle


\section{Experimental measurements}
The experiment poses several challenges. As a photon-hungry technique, resonant inelastic x-ray scattering (RIXS) 
requires a high incident photon flux \cite{kotani2001, ament2011}. Therefore, most RIXS beamlines at synchrotron 
radiations prioritize maximizing photon flux over optimizing beam collimation and size. However, probing an x-ray 
thin-film cavity necessitates grazing incidence, and the angular range of the cavity mode is extremely narrow 
\cite{rohlsberger2004prb}. For instance, the first cavity mode angle of the current sample is approximately 3.5 mrad, 
with an effective angular range of about $90 ~\mu$rad. Therefore, we have to use a highly collimated beam with a 
relatively small size. Focusing optics can lead to a smaller beam size while preserving the total flux; however, this 
comes at the cost of increased beam divergence. Therefore, we used collimated beams by cutting the incident beam with 
slits, and we lose beam intensities as a compromise. 

On the other hand, despite a small incoming beam size, grazing incidence results in large footprints on the sample surface. For example, the grazing incidence will lead to an enlargement factor of $\sim 286$ at 3.5 mrad, i.e., a 50-$\mu$m beam spot will create a 14.3-mm long footprint. As a result, the emission from the sample behaves as a line-like emission source rather than a point-like source. Usually, the source size effect can degrade the energy resolution if the geometry is inappropriate, for example, bringing more challenges to high-resolution Johann and DuMond analyzers. In principle, the emission source shape can be constrained using an aperture, such as a pinhole. However, this would further reduce the detectable emission events. Note that the core-to-core RIXS in the hard x-ray regime features a natural linewidth of several eV, indicating that it is not necessary to use a spectrometer with a meV energy resolution. Additionally, the cross section for the two-step dipole-allowed transition is relatively strong. In this consideration, we choose the von Hamos geometry \cite{hamos1933, wach2020} for the demonstration of cavity effects on RIXS, which can provide energy resolutions ranging from sub-eV to several eV and is compatible with a line-like source.

\subsection{Long footprint} 

\begin{figure}[htbp]
	\centering
	\includegraphics[width=1.0\linewidth]{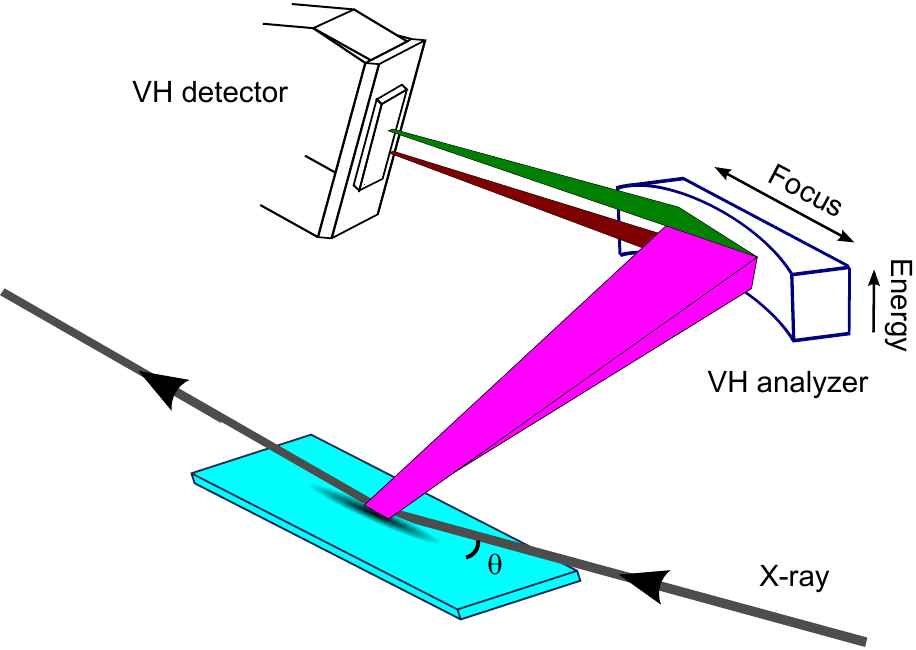}
	\caption{Sketch of the von Hamos spectrometer. The cavity sample was mounted on a high-precision mini hexapod enabling a fine control of the orientation.}
	\label{fig:s0}
\end{figure}

\begin{figure}[htbp]
	\centering
	\includegraphics[width=1.0\linewidth]{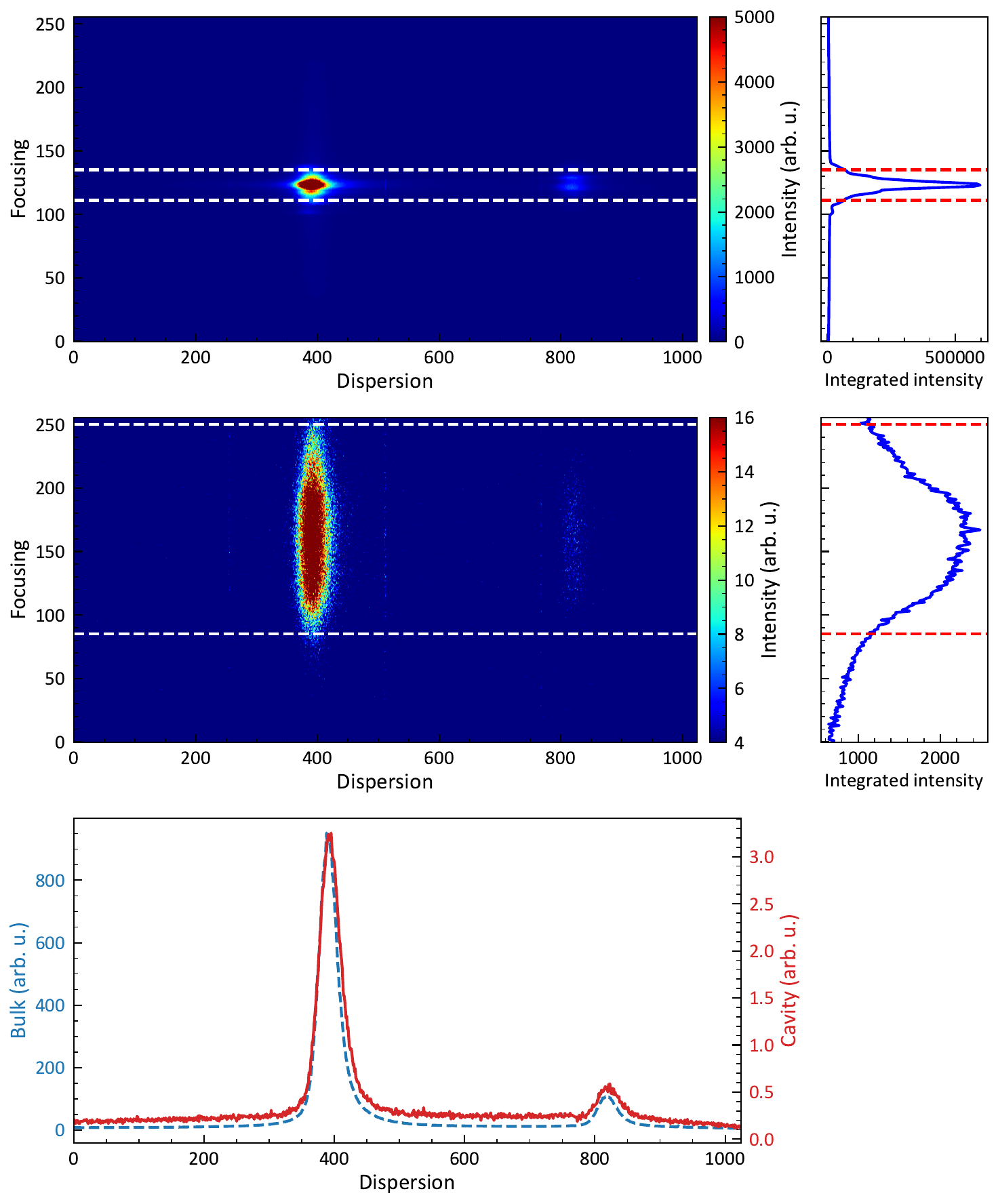}
	\caption{Emission images of the bulk sample (a) and the cavity (b), respectively. The target is much thicker in the bulk sample compared to the cavity, and, thus, the intensities are much higher. (c) The comparison between the emission spectra of bulk and cavity samples.}
	\label{fig:s1}
\end{figure}

The von Hamos spectrometer \cite{hamos1933, wach2020} focuses the emitted x-rays along the incident x-ray footprint and the energy is dispersed over the vertical direction, covering the L$\alpha$ lines. In the experiment, we used eight VH analyzers to increase the solid angle of collection. The analyzed images from all the analyzers are overlapped to increase the signal-to-noise ratio per pixel, taking into account the finite size of the emission detector. Figure \ref{fig:s1} shows the emission images from a bulk sample (operated at a large incident angle, behaving as a point-like source) and the thin-film cavity sample (with a large footprint), respectively. Figure \ref{fig:s1}(a) and (b) use the same incident energy, which is sufficiently higher than the ionization threshold. The emission image focuses onto about 20 pixels for the bulk sample, while it spans over around 160 pixels for the cavity sample at 3.5 mrad. Note that the large footprint slightly affects the energy resolution of the VH spectrometer. Figure \ref{fig:s1}(c) shows a comparison of calibrated emission spectra projected from Fig. \ref{fig:s1}(a) and (b). The spectrum from the cavity sample exhibits a slightly broader profile compared to the bulk sample. The energy resolution for the bulk sample is approximately 1 eV in full width at half maximum as determined from the elastic scattering measurement, while the resolution for the cavity sample is around 2.3 eV, as estimated from the emission spectrum and ray-tracing simulations. 

\subsection{Slit tracing} 
\begin{figure}[htbp]
	\centering
	\includegraphics[width=1.0\linewidth]{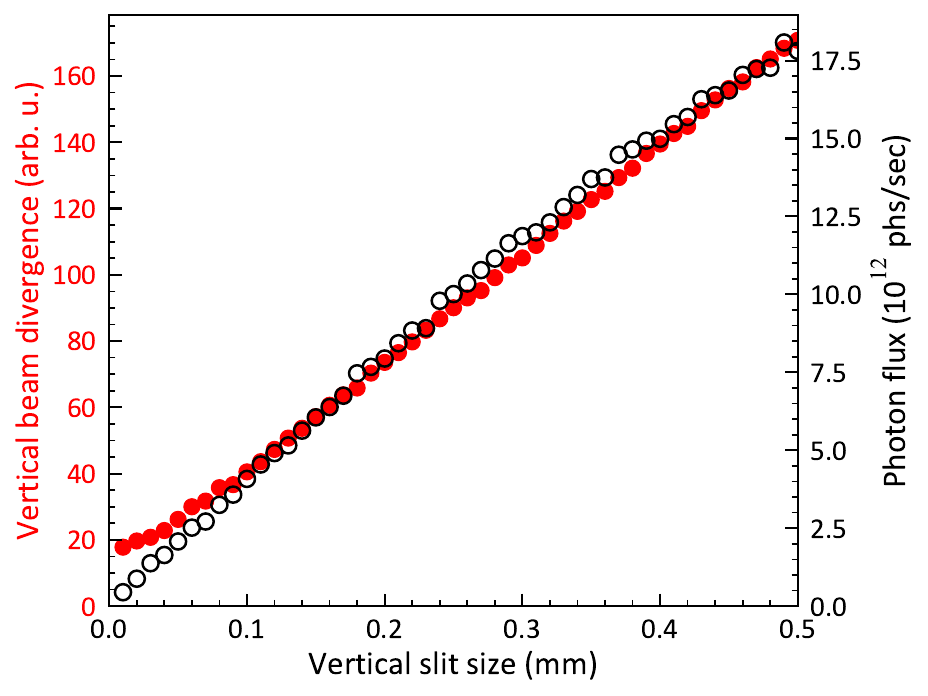}
	\caption{Slit tracing simulation. The beam divergence and photon flux against the slit size. \label{fig:s2}}
\end{figure}

To compromise the beam flux, divergence, and size, we carried out a tracing simulation based on the GALAXIES beamline x-ray source. A strong cavity effect requires a small beam divergence, while RIXS measurement prefers a high photon flux. To achieve a satisfactory beam size and vertical divergence, the incident x-ray beam was cut for the cavity sample using a group of slits. By closing the slit width, the beam divergence can be reduced by sacrificing the incident x-ray beam intensity. Figure \ref{fig:s2} shows the beam divergence and photon flux against the slit size. Compromising the beam divergence and intensity, the slit size was closed to approximately 60 $\mu$m, corresponding to a beam divergence of 30 $\mu$rad during the measurements, and a photon flux of $2.5\times10^{12}$ phs/sec which is almost one-order of magnitude lower than the available maximum flux at GALAXIES beamline.

\subsection{Energy calibration}
Before collecting the emission spectra, we calibrated the detector using the elastic scattering of a bulk sample \cite{zhao2025}. The calibration was performed for an energy range of 8320-8450 eV, covering the L$_{\alpha}$ emission lines. The elastic scattering profiles are fitted using Gaussian functions. The resulted Gaussian centers (pixels) are fitted to a third-order polynomial for scaling the pixels to absolute energies. The energy resolution (approximately 1 eV in full width at half maximum) was also evaluated based on the fit of the elastic scattering profiles.

\section{Sample preparation}
The sample was fabricated using direct-current magnetron sputtering at room temperature \cite{ma2022}. The layers were deposited on a 30$\times$30-mm$^{2}$ superpolished silicon wafer. The structure (i.e., the thickness of each layer), was initially characterized using the grazing incidence x-ray reflectivity (GIXRR) with a Bede D1 diffractometer at the Cu K$\alpha$ ($\sim$ 8.05 keV). However, the angular resolution of the lab source is not high enough to identify the cavity modes, and the cavity was probed at different x-ray energies during the experiment (approximately 10 keV). Therefore, we have to measure the rocking curve at the probe x-ray energy to determine the cavity mode angles for the current measurements.

\subsection{Sample structure}

\begin{figure}[htbp]
	\centering
	\includegraphics[width=\linewidth]{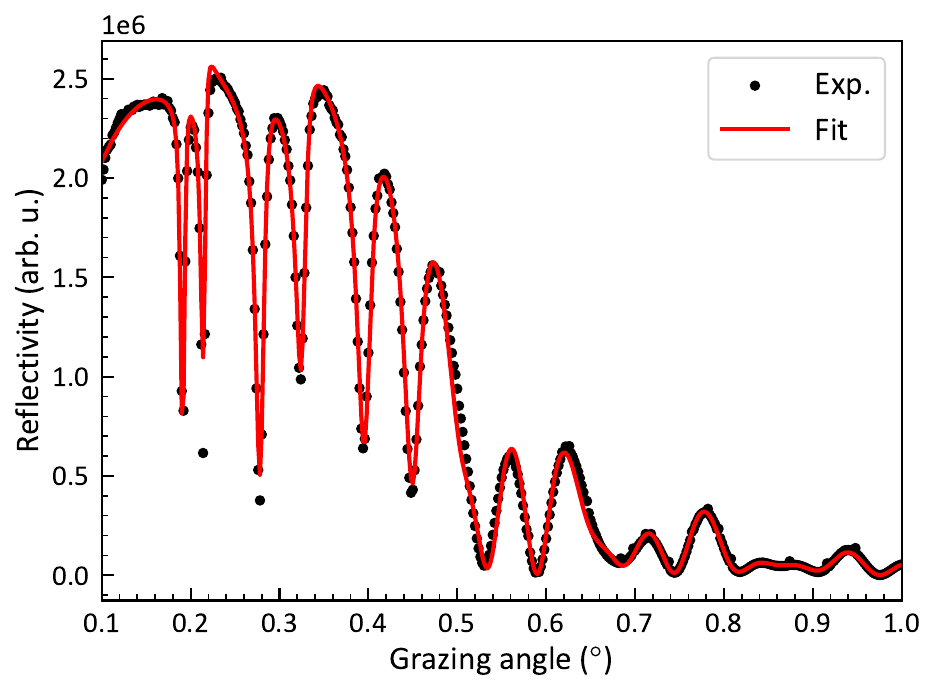}
	\caption{The measured rocking curve (black dotted line) and the fit to data (red solid line).
		\label{fig:s3} }
\end{figure}

Before the RIXS experiment at the GALAXIES beamline, we measured the rocking curve (also known as the x-ray reflectivity, XRR) with an off-resonant x-ray energy at the B16 Test Beamline at the Diamond Light Source, which enabled us to determine the cavity structure by fitting the rocking curve and check the cavity performance. The rocking curve was measured for grazing angles within 0.1\textdegree\ to 1\textdegree, and the fit reasonably reproduces the curve as shown in Fig. \ref{fig:s3}. The fit was performed according to Parratt's algorithm using the GenX software \cite{genx2007} and obtained a cavity structure of 2.4-nm Pt / 20.4-nm C / 2.8-nm WSi$_{2}$ / 20.0-nm C / 14.0-nm Pt deposited on a silicon wafer. Table \ref{tab:s1} lists the fitting parameters, including the thickness, roughness, and density of each layer. The fitted cavity structure is used as the input for the quantum Green's function model to determine the optimized cavity detunings for achieving strong CER and CIS effects. 

\begin{table}[htbp]
    \centering
    \caption{Cavity structure according to the fit in Fig.~\ref{fig:s3}.}
    \label{tab:s1}
    \begin{tabularx}{\linewidth}{l >{\centering\arraybackslash}X c >{\centering\arraybackslash}X c >{\centering\arraybackslash}X c}
        \hline
        Layers & & \thead{Thickness \\ (nm)} & & \thead{Roughness \\ (nm)} & & \thead{Density \\ (g/cm$^3$)} \\
        \hline
         Pt      & & $2.4$  & & $0.23$  & & $21.4$ \\
         C       & & $20.4$ & & $0.10$  & & $2.15$ \\
         WSi$_2$ & & $2.8$  & & $0.49$  & & $9.57$ \\
         C       & & $20.0$ & & $0.10$  & & $2.15$ \\
         Pt      & & $14.0$ & & $0.10$  & & $21.4$ \\
        \hline
    \end{tabularx}
\end{table}

\subsection{Angle identification}

\begin{figure}[htbp]
	\centering
	\includegraphics[width=1.0\linewidth]{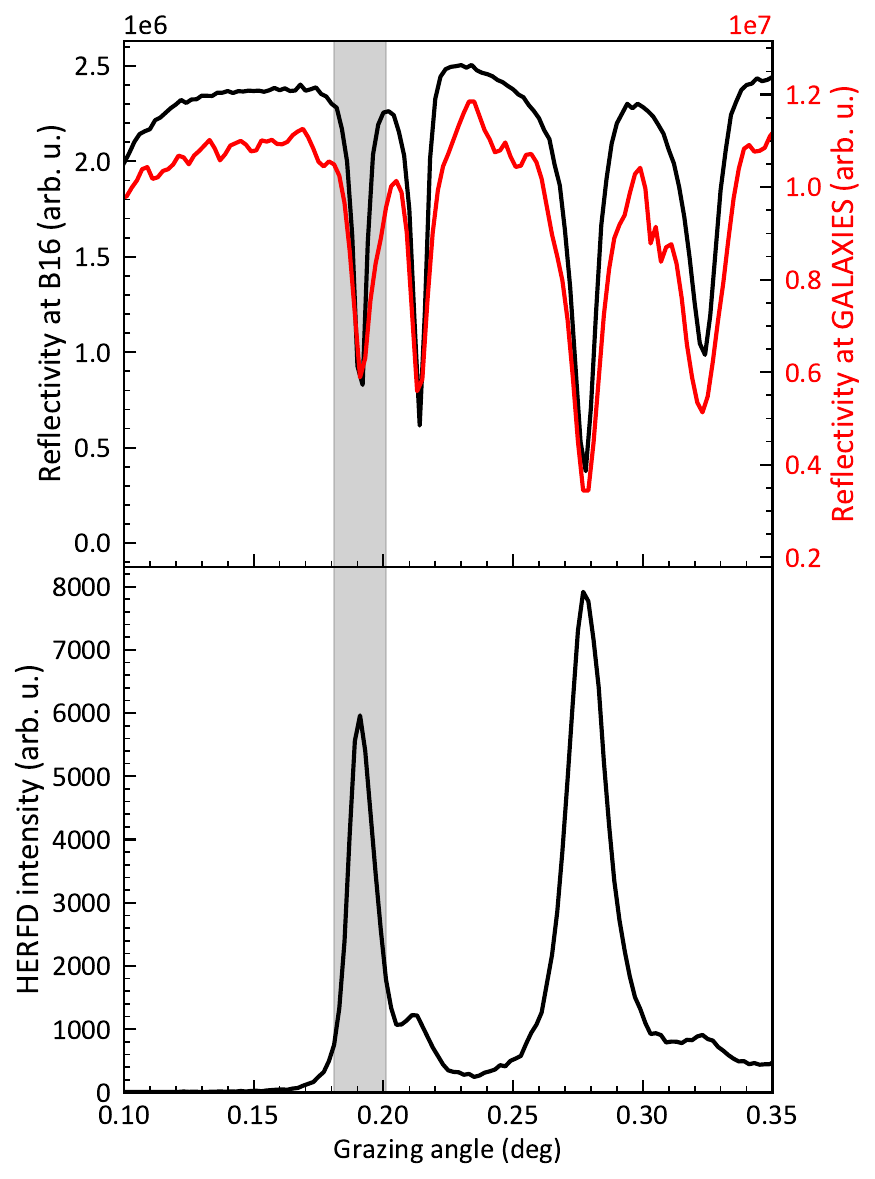}
	\caption{(a) Comparison between the rocking curves measured at the GALAXIES (red curve) and B16 (black curve) beamlines. Note that the curve measured at the GALAXIES beamline is shifted towards lower angles by 0.009\textdegree. (b) The L$_\alpha$ fluorescence intensity as a function of grazing angle, serving as a cross-check of the odd-order cavity mode angles. Note that we apply the same angle shift as done in (a). The shadowed area marks the angle range of 0.18-0.20\textdegree.}
	\label{fig:s7}
\end{figure}

X-ray reflectivity (XRR), i.e. the reflected x-ray intensity against the incident angle, is the key probe to the cavity structure and the cavity mode angles. The dips in the measured rocking curve correspond to the cavity mode angles, where incident x-rays are strongly coupled into the cavity, forming x-ray standing waves. Normally, the rocking curve is measured using a $2\theta$ arm, which is linked to the scan of incident angle, i.e., by scanning the incident angles, the detection arm for the reflected x-rays will be rotated to $2\theta$ in the meantime. However, the 2$\theta$ arm at the GALAXIES beamline is not compatible with the von Hamos spectrometer. To measure the rocking curve, we utilize a position-sensitive detector to monitor the moving reflected x-ray beams during scan of incident angles. However, the direct x-ray beam hinders us from extracting the intensity of reflected x-rays at very small angles, making it difficult to determine the absolute zero angle. On the other hand, the moving reflected beam on the detector also suffers from random hot pixels, therefore, the rocking curve is noisier. The measured rocking curve is compared with the data from the B16 beamline in Fig. \ref{fig:s7}. By shifting the curve towards lower angles by 0.009\textdegree, the cavity mode angles (at the dips) are well reproduced. As the absolute angles are not determined, we use angle offsets throughout the manuscript. On the other hand, the total fluorescence yield detector, i.e., silicon drift detector (SDD) is also used to cross-check the mode angles. The dips in the measured rocking curve correspond to strong standing wave fields, which result in peaks (only for odd orders of cavity mode) in the fluorescence intensities at incident energy above the ionization threshold. This is different from the reflectivity curve which is measured at off-resonant x-ray energies below the ionization threshold. Therefore, the peak positions are not identical to the reflectivity dips. 

\begin{figure}[htbp]
	\centering
	\includegraphics[width=1.0\linewidth]{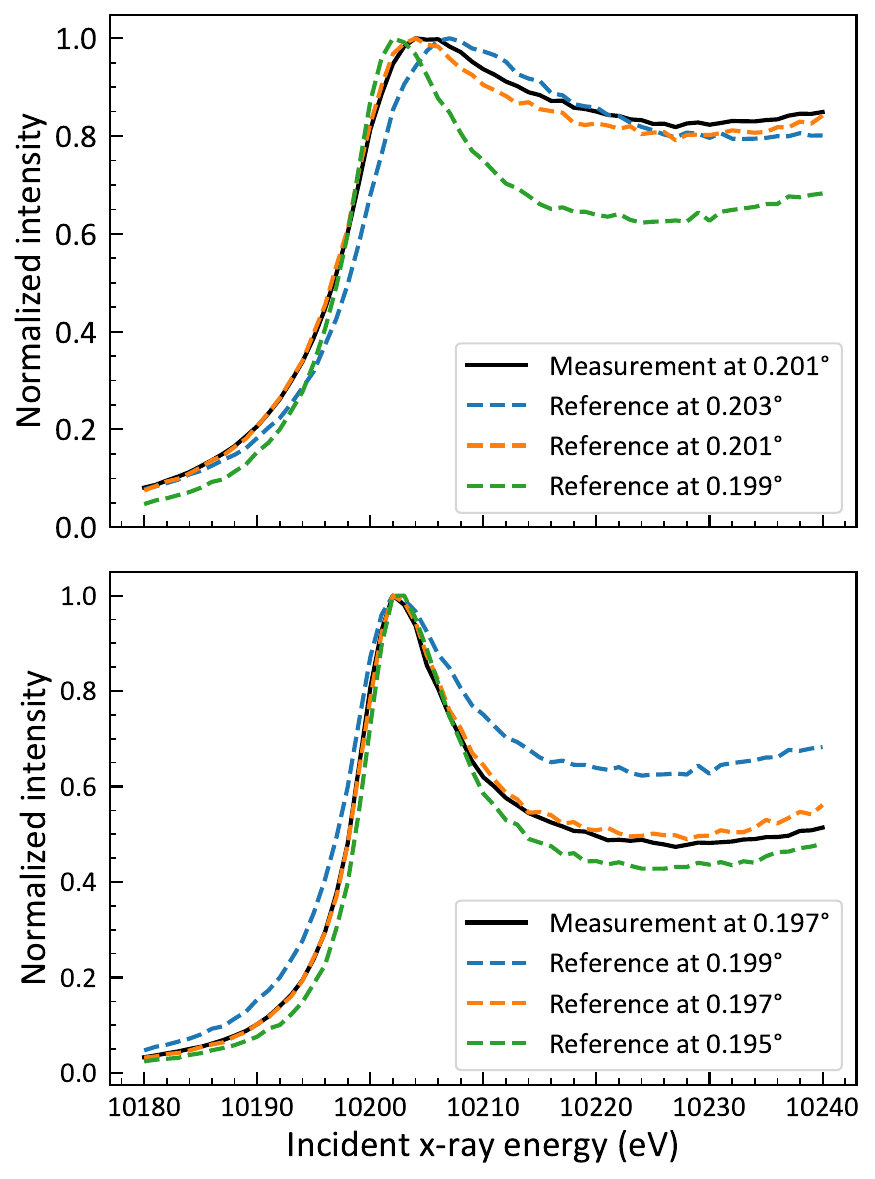}
	\caption{Angle identifications for the RIXS measurements. The fluorescence spectra measured during the RIXS collections are compared with the references.}
	\label{fig:s6}
\end{figure}

The RIXS measurements require a very long collection time, during which the grazing angle might be drifted. During the measurements, the RIXS collection for each incident angle was repeated several times and the SDD recorded the fluorescence spectra in the meantime. The acquisition time of a single collection is limited to a short duration, during which the measurement conditions can be considered stable. On the other hand, we have also collected the reference fluorescence spectra, which are rather sensitive to the incident angles. The reference fluorescence spectra were measured in a relatively short collection time, and, thus, the incident angle can be regarded as constant. Rocking curves were also collected before and after the collection of reference fluorescence spectra to monitor the angle shifts during the measurements. Finally, the incident angles for RIXS are compared to the reference according to the fluorescence spectra. The results in Fig. \ref{fig:s6} indicate a reasonable identification of the incident angles for the RIXS measurements. We estimate that the accuracy of angle identification is better than $10^{-3}$ degrees.

\section{Potential applications}
In addition to the Raman profile discussed in the manuscript, we can extract more information from the measured 2D RIXS plane. For example, we can derive high-energy resolution off-resonant spectroscopy (HEROS) \cite{blachucki2014} by analyzing the off-resonant emission spectrum, the total fluorescence yield x-ray absorption spectra by integrating over emission photon energies, and high-energy resolution fluorescence detection (HERFD) \cite{bauer2014} by integrating a narrow region of emission photon energies. 

\subsection{Total fluorescence yield}
\begin{figure}[htbp]
	\centering
	\includegraphics[width=1.0\linewidth]{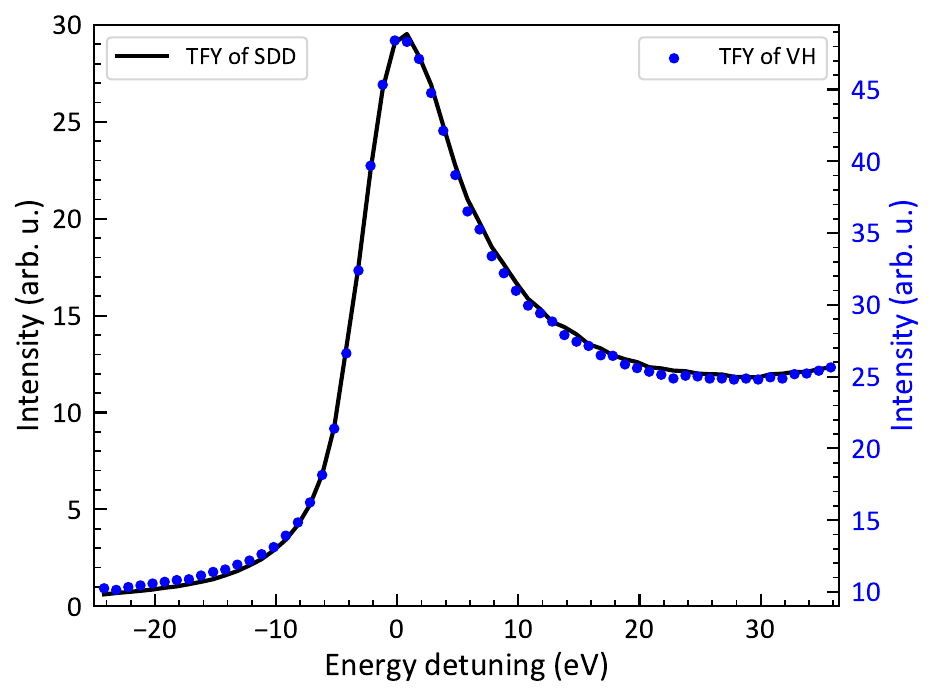}
	\caption{Fluorescence spectra measured using the SDD detector and the VH spectrometer, respectively.
	\label{fig:s4}}
\end{figure}

By integrating over the L$\alpha_1$ emission intensities, we arrive at the x-ray absorption spectra in the sense of total fluorescence yield (TFY), which, in principle, is identical to the fluorescence spectra measured by SDD. Figure \ref{fig:s4} compares the TFY spectra of the SDD and VH spectrometer measured at 8.7 mrad (0.5\textdegree), which show good agreement in the line profile, although they feature different background and noise levels. This comparison is important to verify the von Hamos spectrometer measurement. 

\subsection{CIS induces sharper HERFD-XAS}
\begin{figure}[htbp]
	\centering
	\includegraphics[width=1.0\linewidth]{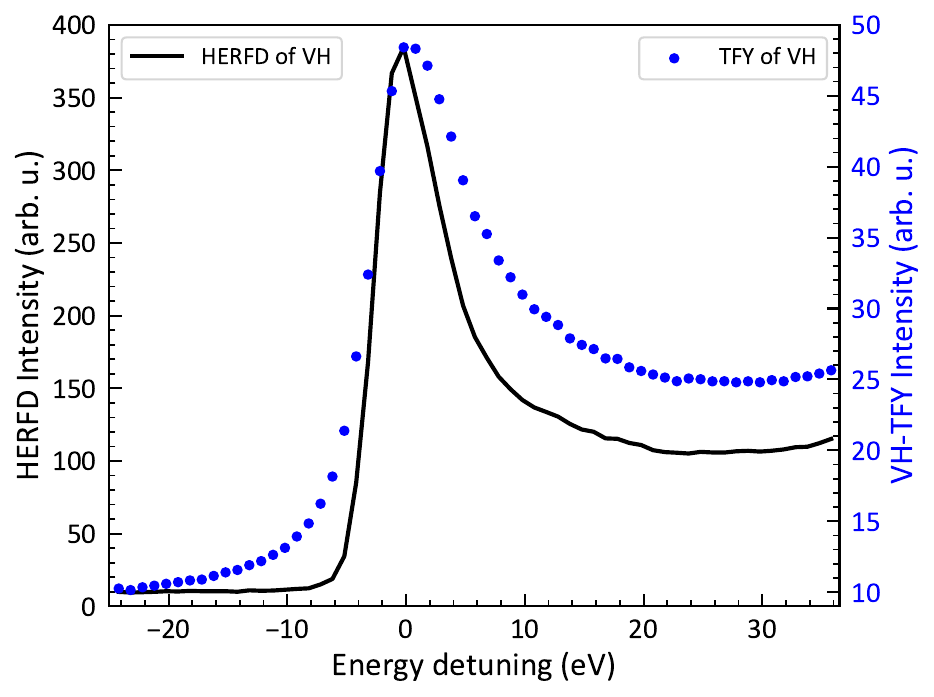}
	\caption{Total fluorescence yield XAS (blue dots) and HERFD-XAS (black line) obtained from the same 2D RIXS plane without cavity effect.}
	\label{fig:s8}
\end{figure}

By integrating over a narrower emission energy band, we arrive at the high-energy-resolution fluorescence detection (HERFD) x-ray absorption spectra (XAS). HERFD-XAS is a modern spectroscopy method that reduces the effect of core-hole lifetime broadening \cite{Hämäläinen1991}. In general, HERFD analyzes the fixed scattering energy at the L$\alpha_1$ peak position, collecting the intensity while scanning the incident energy. Therefore, it avoids integrating the photons along the fixed energy transfer, resulting in sharper features for the bound states. Figure \ref{fig:s8} shows a comparison of TFY-XAS and HERFD-XAS from the same 2D RIXS plane without cavity effect. The HERFD-XAS integrates a range of 2 eV around the L$\alpha_1$ peak, presenting a noticeably sharper feature for the bound state compared to the TFY-XAS. 

\begin{figure}[htbp]
	\centering
	\includegraphics[width=0.9\linewidth]{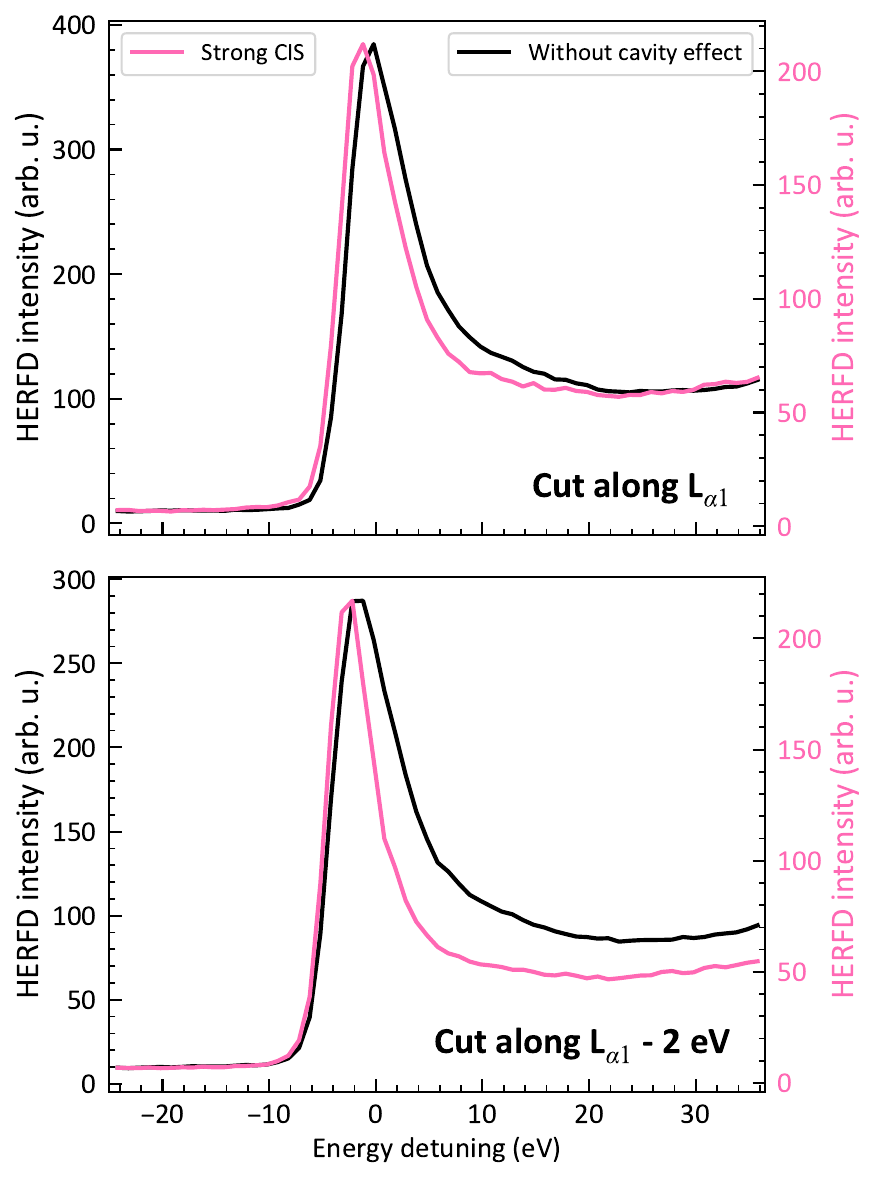}
	\caption{HERFD-XAS spectra for the sample without the cavity effect (black line) or with a strong energy shift effect (pink line). (a) The standard HERFD, where the integration along the emission energy is centered around the L$\alpha_1$ peak, and (b) the center of the integration window is detuned by -2 eV from the L$\alpha_1$ peak.} 
	\label{fig:s9}
\end{figure}

Before the present work, there were no experimental reports about the cavity effect on the HERFD-XAS. Based on the prediction, if a negative cavity-induced energy shift is present, the HERFD integration will not necessarily need to be centered around the L$\alpha_1$ peak, and the feature of the bound state transition could be strengthened. Figure \ref{fig:s9}(a) shows a comparison between two HERFD spectra: one where the cavity effect is absent and one with a large negative cavity-induced energy shift. The white line peak becomes slightly narrower, and the contrast relative to the edge also improves slightly. A more surprising improvement is observed in Fig. \ref{fig:s9}(b), where the integration center is detuned by -2 eV from the L$\alpha_1$ peak. In most scenarios where there is no cavity effect, the HERFD-XAS spectrum is not sensitive to the choice of integration center along the emission energy axis. As shown in the black curves in Fig. \ref{fig:s9}(a) and (b), the profiles of the spectra are very similar, although the detuned spectrum has lower intensity and shifts to negative because its integration center is not aligned with the L$\alpha_1$ emission peak. When a large CIS effect is present, there is a shift toward lower energy detuning on the RIXS plane, which allows for lower emission energy in the HERFD integration without losing intensity for the bound states, however, the intensity of the edge will be reduced. A significantly more pronounced and narrower white line peak is observed in the red curve of Fig. \ref{fig:s9}(b), which still shows strong peak intensity compared to Fig. \ref{fig:s9}(a). It can be seen that the CIS can induce a sharper HERFD spectrum, providing a promising example that cavity-controlled core-to-core RIXS could offer potential applications in modern spectroscopy techniques. Currently, the coherence and brilliance of x-ray source development are ongoing in fourth-generation synchrotrons and free-electron lasers. The combination of x-ray cavity effect and x-ray spectroscopy holds the potential to open up new frontiers and bring new techniques.

\bibliography{SI.Refs.bib}